\documentclass[superscriptaddress,amsmath,amssymb, aps, pra,twocolumn,nofootinbib]{revtex4-2}

\usepackage{graphicx}
\usepackage{dcolumn}
\usepackage{comment}
\usepackage{bm}
\usepackage{mathtools}
\usepackage[colorlinks]{hyperref}
\usepackage{amsthm}
\usepackage[dvipsnames]{xcolor}
 \usepackage{dsfont}
 \usepackage{orcidlink}
 \usepackage{enumerate}
 
\newcommand{\be}{\begin{equation}}
\newcommand{\ee}{\end{equation}}
\newcommand{\beq}{\begin{eqnarray}}
\newcommand{\eeq}{\end{eqnarray}}

\newcommand{\ket}[1]{\mbox{$ | #1 \rangle $}}
\newcommand{\bra}[1]{\mbox{$ \langle #1 | $}}

\newcommand{\ketbra}[1]{\ket{#1}\bra{#1}}
\newcommand{\ketbrac}[1]{\ket{#1}_{\rm c}\bra{#1}}

\newcommand{\Texp}{k_BT/\varepsilon=1.65\pm0.02}

\begin{document}

\title{Correlations in a quantum switch-based heat engine with measurements: \\ A proof-of-principle demonstration} 

\author{Vinicius F. Lisboa}
\affiliation{Center for Natural and Human Sciences, Federal University of ABC, Avenida dos Estados 5001, 09210-580, Santo Andr\'{e}, S\~{a}o Paulo, Brazil}

\author{Pedro R. Dieguez\orcidlink{0000-0002-8286-2645}}\email{dieguez.pr@gmail.com}
\affiliation{International Centre for Theory of Quantum Technologies, University of Gda\'nsk, Jana Ba{\. z}y\'nskiego 8, 80-309 Gda\'nsk, Poland}

\author{Kyrylo Simonov\orcidlink{0000-0001-6764-8555
}}
\affiliation{Fakult\"at f\"ur Mathematik, Universit\"at Wien, Oskar-Morgenstern-Platz 1, 1090 Vienna, Austria}
\affiliation{ICFO - Institut de Ci\`encies Fot\`oniques, The Barcelona Institute of Science and Technology, 08860 Castelldefels, Barcelona, Spain}

\author{Roberto M. Serra\orcidlink{0000-0001-9490-3697}}\email{serra@ufabc.edu.br}
\affiliation{Center for Natural and Human Sciences, Federal University of ABC, Avenida dos Estados 5001, 09210-580, Santo Andr\'{e}, S\~{a}o Paulo, Brazil}

\begin{abstract}
Allowing the order of quantum operations to exist in superposition is known to open new routes for thermodynamic tasks. We investigate a quantum heat engine where energy exchanges are driven by generalized measurements, and the sequence of these operations is coherently controlled in a superposition of causal orders. Our analysis explores how initial correlations between the working medium and the controller affect the engine's performance. Considering uncorrelated, classically correlated, and entangled initial states, we show that entanglement enables the superposed causal order to generate coherence in the working medium, thereby enhancing work extraction and efficiency beyond the separable and uncorrelated cases. Finally, we present a proof-of-principle simulation on the IBM Quantum Experience platform, realizing a quantum switch of two measurement channels with tunable strengths and experimentally confirming the predicted efficiency enhancement enabled by correlation-assisted superposed causal order.

\end{abstract}

\maketitle


\section{Introduction}

Our everyday experience suggests that events occur in a definite order: one event either causes another, follows it, or is causally independent. However, attempts to reconcile quantum theory with the dynamical nature of spacetime in general relativity challenge this view, indicating that causal relations themselves may not be fundamentally predefined. In the absence of a classical spacetime background, even the notions of time evolution or spacelike hypersurfaces on which quantum states are defined become ambiguous \cite{zych2019bell}.

To address this, several abstract frameworks have been developed to describe quantum processes without assuming a definite causal structure~\cite{hardy2005probability, hardy2007towards, chiribella2009beyond, oreshkov2012quantum, chiribella2013quantum, brukner2014quantum, Oeckl_2019}. These formalisms provide a unified operational setting in which indefinite causal order can be formulated and analyzed. Among the physical realizations of this concept, the quantum SWITCH stands out as its most prominent and experimentally demonstrated instance. In this process, a quantum control coherently determines the order in which two operations act on a target system, thereby placing the causal order in superposition~\cite{chiribella2009beyond, chiribella2013quantum}, a feature we refer to as a superposed causal order (SCO). The quantum SWITCH has been realized across a variety of photonic platforms~\cite{procopio2015experimental, rubino2017experimental, goswami2018indefinite, Procopio2019, goswami2020experiments, goswami2020increasing, guo2020experimental, Taddei2021, rubino2021experimental, cao2022quantum, Stroemberg2023, Antesberger2023, Yin2023, rozema2024experimental, wang2025exp, deng2025integrated}, with further proposals involving nuclear magnetic resonance (NMR)~\cite{nie2022experimental} and atom--field interactions~\cite{procopio2025cavity}.

Beyond its foundational implications, indefinite causal order has been shown to enable operational advantages in several areas of quantum technology, including computation~\cite{chiribella2009beyond, chiribella2013quantum, araujo2014computational, Renner2022, EscandonMonardes2023, simonov2024universal, Liu2024, yin2024magic}, communication~\cite{ebler2018enhanced, Procopio2019, goswami2020increasing, Caleffi2020, Mukhopadhyay_2020, Chiribella2021Comm, Bhattacharya2021, Chiribella2021L, Cacciapuoti2022, Das_2022, Caleffi2023, Guha2023, zhao2025communication, wang2025exp}, cryptography~\cite{SpencerWood2025}, metrology, and resource estimation~\cite{zhao2020quantum, Chapeau2021, Bavaresco2021, Chapeau2022, Gao2023, Liu2023, Chapeau2023, Yin2023, Goldberg2023, Mothe2024, weiinc2025, azado2025}, as well as thermodynamics. Thermodynamic applications have likewise benefited from the introduction of SCO, with enhancements demonstrated in Otto engines~\cite{zhao2022influence, xue2025anomalousheatflowquantum}, quantum refrigeration cycles~\cite{felce2020quantum, felce2021refrigeration, cao2022quantum, nie2022experimental, nie2022cooling}, and measurement-based heat-engine and refrigeration schemes implemented in an indefinite causal order~\cite{dieguez2023thermal}. More broadly, SCO-based protocols have been used to optimize diverse thermodynamic tasks, including quantum battery charging, enhancement of extractable work (e.g., in terms of ergotropy \cite{Allahverdyan2004, Francica2017, Bernards2019}), algorithmic cooling, and related processes~\cite{Guha2020, felce2021indefinite, simonov2022work, vieira2023exploring, Goldberg2023thermo, Zhu2023, Simonov2023dice, molitor2024quantum, simonov2025Activation}.

Concurrently, initial correlations between subsystems (whether classical or genuinely quantum) are known to enhance a variety of quantum informational tasks~\cite{RevModPhys.84.1655, PhysRevA.80.044102}. While entanglement is a key resource, more general forms of quantum correlations can also provide advantages in communication~\cite{madhok2013quantum, pirandola2014quantum} and computation~\cite{brodutch2011quantum}. In open quantum systems, understanding how such correlations persist or degrade under thermalization is essential for developing robust, noise-tolerant protocols~\cite{xu2010experimental}. In thermodynamic settings involving coherently controlled thermalization through the quantum SWITCH, initial correlations between the system and the control (particularly entanglement) have been shown to lift the temperature bound for activating work from passive states, enabling ergotropy extraction even at arbitrarily high local temperatures~\cite{simonov2025Activation}. From a complementary perspective, different strategies of work extraction, e.g., local, causally ordered, or cooperative, have been shown to yield distinct ergotropic advantages~\cite{castellano2024local, castellano2025parallel}. Yet, the thermodynamic role of SCO-assisted by initial correlations remains largely unexplored in cyclic heat engine implementations.

Here, we investigate a measurement-driven heat engine powered by a superposition of causal orders and study how initial correlations shape its performance. In particular, we show that classical and quantum correlations between the working medium and the controller extend the operational domain of the engine and enhance its efficiency by activating quantum coherence in the working substance.

Quantum measurements are inherently invasive and can alter a system's internal energy~\cite{yi2017single, elouard2017role, brandner2015coherence, lisboa2022experimental, malavazi2025two}, motivating a series of studies on measurement-powered thermal devices~\cite{campisi2017feedback, chand2017single, chand2017measurement, mohammady2017quantum, elouard2017extracting, chand2018critical, ding2018measurement, elouard2018efficient, buffoni2019quantum, solfanelli2019maximal, jordan2020quantum, behzadi2020quantum, seah2020maxwell, bresque2021two, chand2021finite, lin2021suppressing, PhysRevLett.124.110604, PhysRevLett.121.030604, PRXQuantum.3.020329, PhysRevLett.117.240502, anka2021measurement, alam2022two, manikandan2022efficiently, Obina2022, lisboa2022experimental}. A paradigmatic example is a single-temperature heat engine operating without feedback, where work is extracted via non-selective measurements~\cite{yi2017single}. A proof-of-principle realization using NMR demonstrated such a cycle with two tunable generalized measurement channels: one acting as an effective heat source and the other as an isentropic work-extraction stroke~\cite{behzadi2020quantum, lisboa2022experimental}. When the latter preserves the von Neumann entropy, the associated energy change is identified as work. Experiments have shown that, with suitable tuning of the measurement strengths, these engines can approach unit efficiency while maintaining finite power output~\cite{lisboa2022experimental}.

This model was later generalized to encompass not only heat engines but also refrigerators and thermal accelerators~\cite{dieguez2023thermal}. There, the cycle consists of two generalized measurement channels represented by completely positive trace-preserving (CPTP) maps with adjustable measurement strengths. When these channels are placed in SCO via a quantum SWITCH, work extraction can be activated even for parameters that yield zero efficiency in a definite order~\cite{dieguez2023thermal}.

In this work, we extend that framework by analyzing how initial correlations between the working medium and the controller, ranging from separable correlations to quantum entanglement, affect the operation of a SCO-powered measurement engine. The working substance, initially prepared in a local Gibbs state, undergoes two generalized non-selective measurements in SCO. The second measurement's strength depends on the controller's state, introducing one bit of classical memory that must be erased to close the thermodynamic cycle. Crucially, we show that initial entanglement enables the SCO to generate quantum coherence within the working medium, thereby enhancing the engine's efficiency. We further simulate this process experimentally on a commercial quantum processor, demonstrating the activation of work extraction driven by SCO.

The remainder of the paper is organized as follows. In Section~\ref{sec:model}, we introduce the model of the measurement-driven heat engine with SCO. Section~\ref{sec:results} presents the theoretical analysis of its efficiency and the experimental verification. Finally, in Section~\ref{sec:conclusions}, we draw the conclusions and outlook.


\section{The model} \label{sec:model}

We begin by outlining the model of a heat engine powered by generalized quantum measurements performed in a definite causal order~\cite{lisboa2022experimental}, and then extend it to the case where the order of these measurements is placed in a quantum superposition~\cite{dieguez2023thermal}.

\subsection{Heat engine based on generalized measurements with definite order}\label{sec:HeatEngine}

The fundamental setup of a measurement-based quantum heat engine consists of a working medium, represented by a qubit system $\mathbf{Q}$, a single cold reservoir $\mathbf{R}$ at temperature $T$, and two measurement apparata, $\mathbf{A}$ and $\mathbf{B}$.
The working medium is governed by the Hamiltonian $H = -\varepsilon Z$, where $Z = |0\rangle\langle 0| - |1\rangle\langle 1|$ is the Pauli-$Z$ operator\footnote{Throughout this work, we use the computational basis $\{|0\rangle, |1\rangle\}$, defined by the eigenstates of $H$.}, and the energy gap between the levels is $2\varepsilon$.

The system is initialized in a thermal (Gibbs) equilibrium state,
\begin{equation}\label{eq:iniGibbs}
    \rho^{(0)} = \exp\left[-\beta H\right]/\mathcal{Z}_\beta,
\end{equation}
by fully thermalizing it with the reservoir $\mathbf{R}$, where $\beta=\left(k_{B}T\right)^{-1}$ is the inverse temperature of the qubit, and $\mathcal{Z}_\beta=\operatorname{tr}\left[\exp\left(-\beta H\right)\right]$ is the associated partition function.

The effect of the measurement apparata is modeled as a non-selective measurement using completely positive trace-preserving maps $\mathcal{M}^\lambda[\cdot] = \sum_i M_i^\lambda (\cdot)M_i^{\lambda\dagger}$ with Kraus operators $\{M_i^\lambda\}_{i=1}^4$:
\begin{equation}\label{eq:Kraus}
\begin{matrix*}[l]
M_1^\lambda = \sqrt{\lambda}\ket{0}\bra{0}, & M_2^\lambda = \sqrt{\lambda}\ket{0}\bra{1}, \\
 M_3^\lambda = \sqrt{1-\lambda}\ket{1}\bra{0}, & M_4^\lambda = \sqrt{1-\lambda}\ket{1}\bra{1},
\end{matrix*}
\end{equation}
and tunable strength $0 \leq \lambda \leq 1$. Here we follow the standard convention in measurement-powered engines that, because the Hamiltonian of the working system remains fixed throughout the protocol, isentropic measurement channels are thermodynamically equivalent to work-like strokes, whereas entropy-increasing maps correspond to heat exchange. A generalized measurement channel admits a representation $\mathcal{M}^\lambda[\rho] = \operatorname{Tr}_E[U (\rho \otimes \sigma_E)U^\dagger]$, where $E$ denotes the measurement apparatus (or environment), $\sigma_E$ is its initial state, and $U$ is the unitary interaction between system and apparatus. In this picture, the specific Kraus operators used in \eqref{eq:Kraus} correspond precisely to the map obtained after tracing over uncontrolled degrees of freedom of $E$, which act as an effective reservoir. This interpretation is consistent with microscopic models of measurement-induced thermalization and previous realizations of measurement-powered engines~\cite{behzadi2020quantum, lisboa2022experimental}. Applying this measurement channel leaves the working medium $\mathbf{Q}$ in the state
\begin{equation}\label{eq:MeasChan}
    \mathcal{M}^\lambda[\rho] = \operatorname{diag}[\lambda, 1-\lambda],
\end{equation}
regardless of its initial state $\rho$. Thus, measurements performed by $\mathbf{A}$ and $\mathbf{B}$ are described by the channels $\mathcal{M}^a[\cdot]$ and $\mathcal{M}^b[\cdot]$, with tunable strengths $0 \leq a \leq 1$ and $0 \leq b \leq 1$, respectively.

This setup operates as a three-stroke cycle. In the first two strokes, the medium $\mathbf{Q}$ is sequentially measured by apparatus $\mathbf{A}$, and then by $\mathbf{B}$. Since each apparatus implements a measurement channel of the form given in \eqref{eq:MeasChan}, these strokes transform the state of $\mathbf{Q}$ as follows:
\begin{eqnarray}
    \rho^{(1)} &=& \mathcal{M}^a[\rho^{(0)}] = \operatorname{diag}[a, 1-a], \label{eq:state1Stroke} \\
    \rho^{(2)} &=& \mathcal{M}^b[\rho^{(1)}] = \operatorname{diag}[b, 1-b]. \label{eq:state2Stroke}
\end{eqnarray}
The third and last stroke consists in the thermalization of the medium $\mathbf{Q}$ with the cold reservoir $\mathbf{R}$, thereby completing the cycle: $\rho^{(3)} = \rho^{(0)}$. Each stroke modifies the internal energy $U^{(\ell)} = \operatorname{tr}[H\rho^{(\ell)}]$ and entropy $S^{(\ell)} = -\operatorname{tr}[\rho^{(\ell)} \ln \rho^{(\ell)}]$ of the working medium, where $\ell \in \{1,2,3\}$.

By adjusting the measurement strengths $a$ and $b$, one can control the exchange of internal energy $\Delta U^{(\ell)} = U^{(\ell)} - U^{(\ell - 1)}$, which can be interpreted as either heat or work depending on the associated entropy variation $\Delta S^{(\ell)} = S^{(\ell)} - S^{(\ell - 1)}$. Specifically, if $\Delta S^{(\ell)} \neq 0$, the exchanged energy $\mathcal{Q}^{(\ell)} = \Delta U^{(\ell)}$ is classified as heat, whereas $\Delta S^{(\ell)} = 0$ implies that the exchanged energy corresponds to extracted ($\mathcal{W}^{(\ell)} \geq 0$) or invested ($\mathcal{W}^{(\ell)} \leq 0$) work, given by $\mathcal{W}^{(\ell)} = -\Delta U^{(\ell)}$.

The device functions as a heat engine when:
\begin{enumerate}[1)]
\item the first stroke provides heat $\mathcal{Q}^{(1)} := \mathcal{Q}_{\rm hot} \ge 0$ from $\mathbf{A}$ to $\mathbf{Q}$ (acting as a hot reservoir),
\item the second stroke is isentropic and extracts work $\mathcal{W}^{(2)} := \mathcal{W}_{\rm ext} \ge 0$, and
\item the third stroke dissipates heat $\mathcal{Q}^{(3)} := \mathcal{Q}_{\rm cold} \le 0$ back to $\mathbf{R}$.
\end{enumerate}
Computing the energy and entropy variations for each stroke (see Appendix~\ref{app:heatEngVar}), we find that the extracted work is given by
\begin{eqnarray}
    \mathcal{W}_{\rm ext} &=& \varepsilon \left( 1 - 2a + |1 - 2a| \right)
\end{eqnarray}
provided that the measurement strengths $a$ and $b$ satisfy
\begin{eqnarray}\label{eq:heatEngCond}
    \frac{1}{2}\left(1-\tanh(\beta \varepsilon)\right) < a &<& \frac{1}{2}\left(1+\tanh(\beta \varepsilon)\right), \\
    b &=& \frac{1}{2}\left( 1 + |1-2a| \right). \label{eq:strengthB}
\end{eqnarray}
Under these conditions, the efficiency of the heat engine is (see Table~\ref{tab:energy1} in Appendix~\ref{app:heatEngVar} summarizing each stroke),
\begin{equation}\label{eq:eta}
    \eta = \frac{\mathcal{W}_{\rm ext}}{\mathcal{Q}_{\rm hot}} = \frac{1-2a + |1-2a|}{1-2a+\tanh(\beta\varepsilon)}.
\end{equation}
A non-zero extracted work $\mathcal{W}_{\rm ext} > 0$ and, consequently, a non-zero efficiency \eqref{eq:eta} are achieved under a stricter constraint on $a$,
\begin{equation}
    \frac{1}{2}\left(1-\tanh(\beta \varepsilon)\right) < a < \frac{1}{2}.
\end{equation}
In this case, the efficiency simplifies to
\begin{equation}
    \eta_{>0} = 2 \left(1+\frac{\tanh(\beta \varepsilon)}{1-2a} \right)^{-1}.
\end{equation}

\begin{figure*}[t!]
    \centering
    \includegraphics[width=2.1\columnwidth]{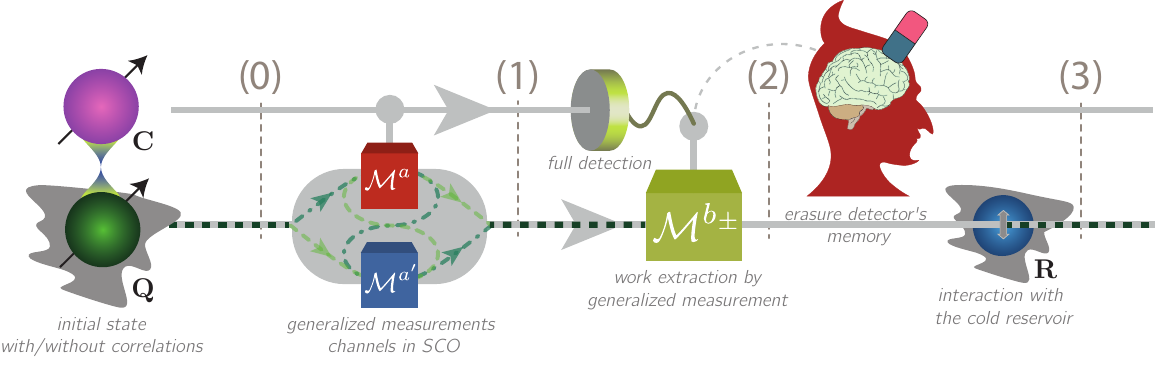}
    \caption{\textbf{Heat engine protocol powered by a quantum SWITCH on two measurement channels.} \textbf{(0)} The protocol starts with the medium $\mathbf{Q}$ and controller $\mathbf{C}$ sharing thermal correlations, where $\mathbf{Q}$ is locally in a thermal state as given by \eqref{eq:localThermalState}. \textbf{(0-1)} In the first stroke, generalized measurements are performed on $\mathbf{Q}$ via measurement apparata $\mathbf{A}$ and $\mathbf{A'}$, with their causal order controlled by the quantum SWITCH. \textbf{(1-2)} To enable work extraction in the second stroke, a generalized measurement is applied to $\mathbf{Q}$ using apparatus $\mathbf{B}$, conditioned on the outcome of a prior measurement of the controller $\mathbf{C}$ of the quantum SWITCH.
    \textbf{(2-3)} In the final stroke, the working medium $\mathbf{Q}$ is thermalized with the reservoir $\mathbf{R}$, and the memory of the 1-bit meter used to measure the state of $\mathbf{C}$ is erased. The measurement outcome is discarded if the corresponding heat engine conditions \eqref{eq:heatEngCondSwitch1}-\eqref{eq:heatEngCondSwitch3} are not satisfied.
    }
    \label{fig:protocol}
\end{figure*}

\subsection{Extended heat engine with a SCO of generalized measurements}\label{sec:HeatEngineSWITCH}

The measurement-based heat engine described above can be extended to operate under SCO of measurements, which has been shown to enhance its performance~\cite{dieguez2023thermal}. This is realized by introducing an additional measurement apparatus, $\mathbf{A}'$, during the first stroke, characterized by a measurement strength $a'$. In this setting, heat exchange occurs between the working medium $\mathbf{Q}$ and both devices, $\mathbf{A}$ and $\mathbf{A}'$. The corresponding generalized measurement channels, $\mathcal{M}^a[\cdot]$ and $\mathcal{M}^{a'}[\cdot]$, are applied through a quantum SWITCH, a higher-order operation that coherently superposes their two possible causal sequences.

Formally, the quantum SWITCH acts as a supermap combining $\mathcal{M}^a$ and $\mathcal{M}^{a'}$ into a new map,
\begin{equation}
    \mathcal{S}(\mathcal{M}^{a}, \mathcal{M}^{a'})[\varrho_{\rm in}] = \sum_{i,j=1}^4 K_{ij} \varrho_{\rm in} K_{ij}^\dagger,
\end{equation}
where the Kraus operators $\{K_{ij}\}_{i,j=1}^4$ are given by
\begin{equation}
    K_{ij} = M_j^{a'} M_i^a \otimes |0\rangle_{\rm c}\langle 0| + M_i^a M_j^{a'} \otimes |1\rangle_{\rm c}\langle 1|,
\end{equation}
and $\{|0\rangle_{\rm c}, |1\rangle_{\rm c}\}$ denotes the computational basis of the control qubit $\mathbf{C}$ that determines the causal order. The state $\varrho_{\rm in}$ represents the joint initial state of the working medium and the control system.

Physically, the quantum SWITCH coherently controls whether $\mathcal{M}^a$ precedes $\mathcal{M}^{a'}$, or vice versa, effectively placing these operations in a superposition of causal orders. Such a process can be implemented with present-day technologies, for instance, in photonic platforms using Mach--Zehnder or Sagnac interferometers, or simulated on cloud-based quantum processors such as IBM Quantum Experience.

The complete extended model of the heat engine is depicted in Fig.~\ref{fig:protocol}. The process begins with the working medium $\mathbf{Q}$ and the controller $\mathbf{C}$ sharing an initial joint state $\varrho_{\rm in}$, which is locally thermal with respect to the cold reservoir $\mathbf{R}$,
\begin{equation}\label{eq:localThermalState}
    \operatorname{tr}_{\rm c}[\varrho_{\rm in}] = \rho^{(0)},
\end{equation}
where the index $c$ denotes the control degree of freedom associated with $\mathbf{C}$. In the first stroke, the state of $\mathbf{Q}$ is measured by the apparata $\mathbf{A}$ and $\mathbf{A}'$, whose causal order is coherently controlled via the quantum SWITCH. Unlike in the model introduced in Section~\ref{sec:HeatEngine}, the state $\rho^{(1)}$ of the working medium may acquire coherence, whose amount depends on the operation mode of the SWITCH, the \textit{incoherent} and \textit{coherent} modes.

In the incoherent mode, the controller $\mathbf{C}$ remains unmeasured, so that the SWITCH implements a probabilistic mixture of the two possible causal orders of~$\mathbf{A}$ and~$\mathbf{A}'$. The working medium is then left in the state
\begin{eqnarray}\label{eq:firstStrokeStateSwitchIncoh}
    \rho^{(1)}_{\rm inc} &=& \operatorname{tr}_{c}[\mathcal{S}(\mathcal{M}^{a}, \mathcal{M}^{a'})[\varrho_{\rm in}]] \nonumber \\
    &=& \operatorname{diag}[\bar{a}^{\rm inc}, 1-\bar{a}^{\rm inc}],
\end{eqnarray}
where
\begin{equation}\label{eq:incStrength}
    \bar{a}^{\rm inc} = \lambda[\varrho_{\rm in}] a' + (1 - \lambda[\varrho_{\rm in}]) a,
\end{equation}
with $\lambda[\varrho_{\rm in}] \in [0, 1]$ determined by the initial joint state of $\mathbf{C}$ and $\mathbf{Q}$, and $\bar{a}^{\rm inc} \in [\min(a, a'), \max(a, a')]$. Hence, in the incoherent mode, the state of $\mathbf{Q}$ after the first stroke remains diagonal in the energy basis and is completely characterized by the effective measurement strength $\bar{a}^{\rm inc}$, independently of any off-diagonal structure present in the joint initial state. Therefore, the first and subsequent strokes reproduce the dynamics of the heat engine introduced in Section~\ref{sec:HeatEngine}, yielding an efficiency
\begin{equation}\label{eq:etaIncoh}
    \eta^{\rm inc} = \frac{1-2\bar{a}^{\rm inc} + |1-2\bar{a}^{\rm inc}|}{1-2\bar{a}^{\rm inc} + \tanh(\beta\varepsilon)}.
\end{equation}

In the coherent mode, the controller $\mathbf{C}$ is measured in the orthonormal basis $\{|\pm\rangle_{\rm c}\}$, revealing one of two outcomes and leaving $\mathbf{Q}$ in one of the corresponding states
\begin{eqnarray}
    \rho^{(1)}_{\pm} &=& \frac{1}{p^{(\pm)}}\operatorname{tr}_{c}[\mathcal{S}(\mathcal{M}^{a}, \mathcal{M}^{a'})[\varrho_{\rm in}](\mathbb{I} \otimes |\pm\rangle_{\rm c}\langle\pm|)] \nonumber \\
    &=& \begin{pmatrix} \bar{a}_{\pm} & \rho_{01, \pm}^{(1)} \\ \rho_{01, \pm}^{(1)*} & 1-\bar{a}_{\pm}\end{pmatrix}, \label{eq:firstStrokeStateSwitch}
\end{eqnarray}
with probability $p^{(\pm)} = \operatorname{tr}[\mathcal{S}(\mathcal{M}^{a}, \mathcal{M}^{a'})[\varrho_{\rm in}](\mathbb{I} \otimes |\pm\rangle_{\rm c}\langle\pm|)]$, as detailed in Appendix~\ref{app:Switch}. In general, the coherences $\rho_{01,\pm}^{(1)}$ depend on the off-diagonal blocks of $\varrho_{\rm in}$, so that initial correlations between $\mathbf{Q}$ and $\mathbf{C}$ can generate coherence in the state of the working medium only in the coherent mode. We quantify coherence using its $l_{1}$-norm, $C_{l_{1}}(\rho)=\sum_{i\neq j}|\rho_{ij}|$~\cite{baumgratz2014quantifying}, since this measure directly captures the off-diagonal terms generated in the coherent mode of the quantum SWITCH and is sufficient for assessing whether coherence enhances the thermodynamic performance of the engine.

In the second stroke, the working medium $\mathbf{Q}$ is measured by the apparatus $\mathbf{B}$ with measurement strength $b$, chosen such that the isentropicity condition $\Delta S^{(2)} = 0$ is satisfied. A careful treatment of the states produced during the first stroke is required because: (i) the conditional states in \eqref{eq:firstStrokeStateSwitch} may contain non-zero off-diagonal elements~$\rho_{01,\pm}^{(1)}$, which contribute to their entropy, and (ii) these states are not necessarily identical for the two measurement outcomes of $\mathbf{C}$. Consequently, the required strength depends on the measurement outcome of $\mathbf{C}$, $b \equiv b_\pm$, and so do the corresponding states $\rho^{(\ell)} \equiv \rho^{(\ell)}_\pm$\footnote{In this notation, we assume $\rho_{+}^{(0)} = \rho_{-}^{(0)} = \rho^{(0)}$.}, produced in each stroke.

Defining the conditional internal energies $U^{(\ell)}_{\pm} = \operatorname{tr}[H\rho^{(\ell)}_{\pm}]$ and entropies $S^{(\ell)}_{\pm} = -\operatorname{tr}[\rho^{(\ell)}_{\pm}\ln\rho^{(\ell)}_{\pm}]$, together with their variations $\Delta U^{(\ell)}_{\pm} = U^{(\ell)}_{\pm} - U^{(\ell-1)}_{\pm}$ and $\Delta S^{(\ell)}_{\pm} = S^{(\ell)}_{\pm} - S^{(\ell-1)}_{\pm}$, we impose the thermodynamic conditions $\mathcal{Q}_{\rm hot}^{(\pm)} = \Delta U_{\pm}^{(1)} \geq 0$ in the first stroke and $\mathcal{W}_{\rm ext}^{(\pm)} = -\Delta U_{\pm}^{(2)} \geq 0$ in the second. Under these requirements, the extractable work reads
\begin{equation}\label{eq:workPostSelect}
    \mathcal{W}_{\rm ext}^{(\pm)} = \varepsilon\Bigl(1-2\bar{a}_{\pm} + \sqrt{(1-2\bar{a}_{\pm})^2 + 4|\rho_{01, \pm}^{(1)}|^2}\Bigr),
\end{equation}
which can be obtained from $\mathbf{Q}$ by setting
\begin{equation}
    b_\pm = \frac{1}{2}\left( 1 + \sqrt{(1-2\bar{a}_{\pm})^2 + 4|\rho_{01, \pm}^{(1)}|^2} \right),
\end{equation}
subject to the constraints
\begin{eqnarray}
    \label{eq:heatEngCondSwitch1} \bar{a}_{\pm} &>& \frac{1}{2}\left(1-\tanh(\beta \varepsilon)\right), \\
    \label{eq:heatEngCondSwitch2} \bar{a}_{\pm} &<& \frac{1}{2}\left(1+\tanh(\beta \varepsilon)\right), \\
    \label{eq:heatEngCondSwitch3} |\rho_{01, \pm}^{(1)}| &<& \frac{1}{2}\sqrt{\tanh^2(\beta \varepsilon) - (1-2\bar{a}_{\pm})^2},
\end{eqnarray}
which depend on whether the measurement of $\mathbf{C}$ yields the outcome $|+\rangle_{\rm c}$ or~$|-\rangle_{\rm c}$ (see Appendix~\ref{app:varEnerEntrSwitch}). Notice that \eqref{eq:workPostSelect} depends quadratically on $\sqrt{2}|\rho_{01,\pm}^{(1)}|$, which coincides with the purity-based contribution of coherence associated with these off-diagonal elements. Thus, although we track coherence via the $l_1$-norm for clarity, the quantity that directly enters the thermodynamic expressions is precisely the purity-based coherence generated during the first stroke. A more detailed analysis based on the relative entropy of coherence (naturally linked to entropy production and irreversibility) is beyond the scope of the present work and will be explored in future investigations.

Importantly, and in contrast to the heat engine model of Section~\ref{sec:HeatEngine}, the extracted work $\mathcal{W}_{\rm ext}^{(\pm)}$ remains strictly positive for all $\bar{a}_{\pm}$ satisfying \eqref{eq:heatEngCondSwitch1}--\eqref{eq:heatEngCondSwitch2}, provided that $\rho_{01, \pm}^{(1)} \neq 0$. Finally, enforcing $\mathcal{Q}_{\rm cold}^{(\pm)} = \Delta U_{\pm}^{(3)} \leq 0$ in the last stroke leads the working medium to rethermalize with the reservoir $\mathbf{R}$, restoring its initial local state, $\rho^{(3)}_\pm = \rho^{(0)}$.

Before discussing the efficiency of the extended heat engine, it is necessary to address a subtle point that arises in the coherent mode of the quantum SWITCH. The constraints \eqref{eq:heatEngCondSwitch1}--\eqref{eq:heatEngCondSwitch3} may not be simultaneously satisfied for both conditional states in \eqref{eq:firstStrokeStateSwitch}. When this occurs, the measurement outcome of $\mathbf{C}$ associated with a non-compliant state must be discarded. Together with the explicit dependence of each stroke on the detected outcome, this raises the question of the thermodynamic cost of erasing the information gained from $\mathbf{C}$.

To account for this, we model the detector's memory as a classical register that becomes perfectly correlated with the control state after readout. The erasure of this register follows Landauer's principle~\cite{Bennett2003, berut2012experimental, Reeb2014, Goold2015, Chattopadhyay2025}, which prescribes a minimal work cost when the process is carried out quasistatically. Assuming that the durations of readout, feedback, and erasure are negligible and that any amplification costs are ignored~\cite{elouard2017extracting}, the minimal erasure cost is
\begin{equation}\label{eq:workCost}
    \mathcal{W}_{\rm cost} = -\beta^{-1}_{\rm D} \ln(2),
\end{equation}
where $\beta_{\rm D} = (k_B T_{\rm D})^{-1}$ is the inverse temperature of the detector, and $k_B$ is Boltzmann's constant. In line with standard analyses of information engines, we assume that the amplification of the microscopic detector signal has negligible energetic cost, and that the detector memory is rethermalized with a reservoir at temperature $T_{\rm D}$ after each cycle, ensuring that no entropy accumulates in the memory and that the engine operates in a stable cyclic regime.

To compare the performance of the two branches of the coherent control $|\pm\rangle_{\rm c}$, we distinguish between the work-extracting branch and the non--work-extracting branch, while keeping the global protocol deterministic. In the desired branch (e.g., $|-\rangle_{\rm c}$), we apply the isentropic channel $\mathcal{M}^{b_-}$ that maximizes the work extraction. In the complementary branch, we do not post-select the outcome. Instead, we apply the fixed measurement $\mathcal{M}^b$ with strength $b = (1+\tanh(\beta \varepsilon))/2$, which deterministically returns the working medium to the thermal state $\rho^{(0)}$. This map is not isentropic and satisfies $\mathcal{Q}_{\mathrm{hot}}^{(2)} = - \mathcal{Q}_{\mathrm{hot}}^{(1)} < 0$, thereby ensuring that no work is extracted and no heat is exchanged with the cold reservoir. Importantly, after this operation both branches end in the same local state $\rho^{(0)}$, so the engine remains cyclic without any post-selection.

Under these conditions, and in the limit of a low-temperature detector $T_{\rm D}$ such that $\mathcal{W}_{\rm cost}$ is negligible, the performance of the desired branch is characterized by the efficiency
\begin{equation}\label{eq:effPS}
    \begin{aligned}
        \eta^{(\pm)}&:=\frac{\mathcal{W}^{(\pm)}_{\rm ext}}{\mathcal{Q}^{(\pm)}_{\rm hot}}\\
        &\;=\frac{1 - 2\bar{a}_{\pm} + \sqrt{(1 - 2\bar{a}_{\pm})^2 + 4|\rho_{01, \pm}^{(1)}|^2}}{1 - 2\bar{a}_{\pm} + \tanh(\beta \varepsilon)},
    \end{aligned}
\end{equation}
where, for each outcome, the corresponding pair $(\bar{a}_{\pm}, \rho_{01,\pm}^{(1)})$ individually satisfies the heat engine constraints \eqref{eq:heatEngCondSwitch1}--\eqref{eq:heatEngCondSwitch3}, without requiring that both do so simultaneously.

If, instead, both outcomes are retained, the total supplied heat in the coherent mode coincides with that in the incoherent mode,
\begin{eqnarray}
    \langle \mathcal{Q}^{\text{coh}}_{\rm hot} \rangle &=& p^{(+)} \langle \mathcal{Q}_{\rm hot}^{(+)} \rangle + p^{(-)} \langle \mathcal{Q}_{\rm hot}^{(-)} \rangle \nonumber \\
    &=& \mathcal{Q}^{\rm inc}_{\rm hot},
\end{eqnarray}
where the superscript ``coh'' denotes the coherent mode of the quantum SWITCH, wherein an isentropic channel is always performed in stroke 2. The corresponding average extractable work in the second stroke is
\begin{equation}\label{eq:extWorkSwitch}
    \langle \mathcal{W}_{\rm ext}^{\rm coh} \rangle = p^{(+)} \mathcal{W}_{\rm ext}^{(+)} + p^{(-)} \mathcal{W}_{\rm ext}^{(-)}.
\end{equation}
The resulting efficiency of the extended engine in the coherent mode (see Table~\ref{tab:energy2} in Appendix~\ref{app:varEnerEntrSwitch}, which summarizes each outcome-conditioned stroke) is then
\begin{equation}\label{eq:etaSwitch}
    \eta^{\rm coh} = \frac{\langle\mathcal{W}_{\rm ext}^{\rm coh}\rangle + \mathcal{W}_{\rm cost}}{\langle\mathcal{Q}_{\rm hot}^{\rm coh}\rangle} = \underbrace{\eta^{\rm inc} + \Delta \eta^{\rm coh}}_{:= \tilde{\eta}^{\rm coh}} - \eta_{\rm cost},
\end{equation}
where
\begin{eqnarray} \label{eq:delta_eta_coh}
    \Delta \eta^{\rm coh} &=& \frac{\sum_{\pm} p^{(\pm)} \sqrt{(1-2\bar{a}_{\pm})^2 + 4|\rho_{01, \pm}^{(1)}|^2} - |1-2\bar{a}^{\rm inc}|}{1-2\bar{a}^{\rm inc}+\tanh(\beta\varepsilon)}, \nonumber \\
    \eta_{\rm cost} &=& \frac{\beta^{-1}_{\rm D} \ln(2)}{1-2\bar{a}^{\rm inc}+\tanh(\beta\varepsilon)},
\end{eqnarray}
and both parameter pairs $(\bar{a}_{\pm}, \rho_{01,\pm}^{(1)})$ must satisfy \eqref{eq:heatEngCondSwitch1}--\eqref{eq:heatEngCondSwitch3} simultaneously. Here, $\Delta \eta^{\rm coh}$ quantifies the positive contribution from the SCO (see Appendix~\ref{app:varEnerEntrSwitch}), whereas $\eta_{\rm cost}$ captures the negative correction associated with memory erasure.

The coherent mode of the quantum SWITCH outperforms the incoherent one whenever $\Delta \eta^{\rm coh} > \eta_{\rm cost}$. Since $\eta_{\rm cost}$ depends on the detector temperature, this condition defines a critical threshold
\begin{equation}
    T_{\rm D}^{\rm crit}=\frac{\sum_\pm p^{(\pm)}\sqrt{(1-2\bar{a}_{\pm})^2 + 4|\rho_{01, \pm}^{(1)}|^2} - |1-2\bar{a}^{\rm inc}|}{k_B \ln(2)},
\end{equation}
below which the coherent mode achieves a performance advantage.


\section{Results}\label{sec:results}


\subsection{Enhancement of work by initial correlations} \label{subsec:theor}

\begin{figure*}[t!]
    \centering
    \includegraphics[width=\textwidth]{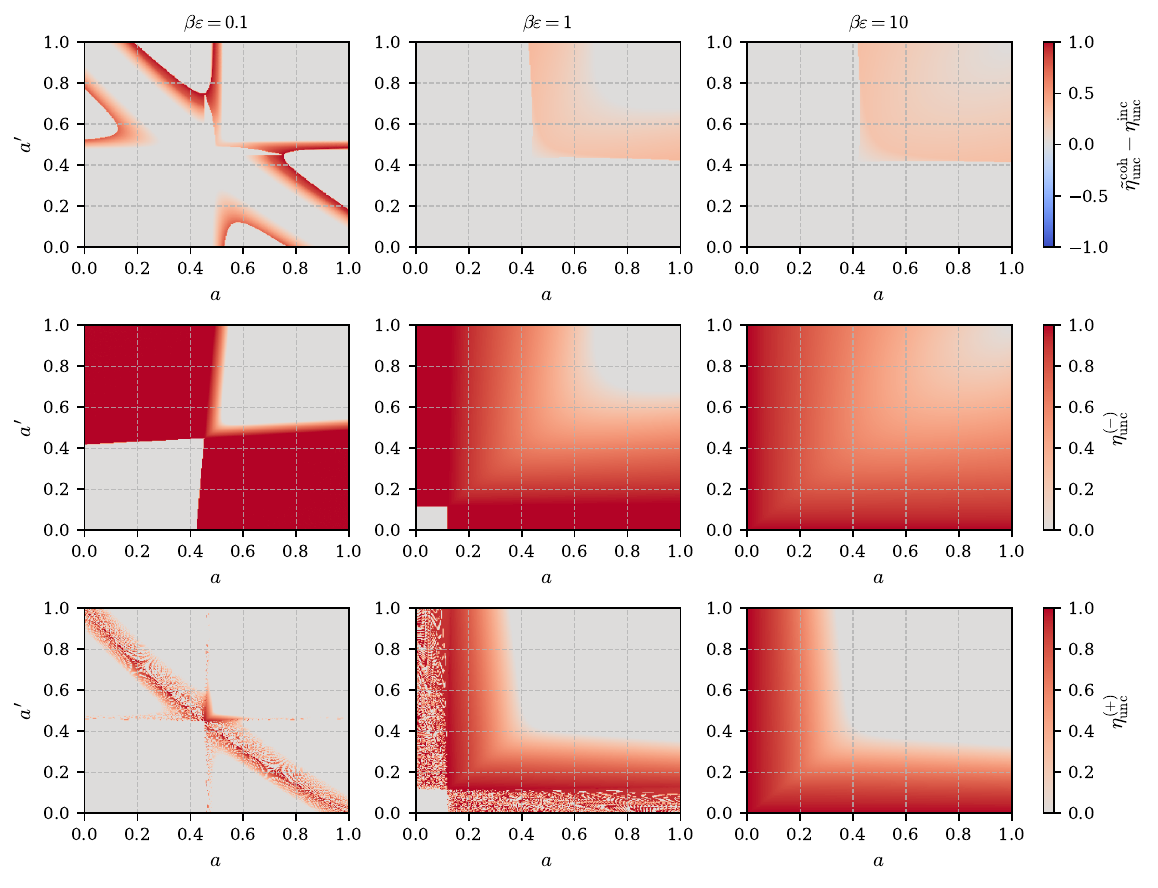}
    \caption{\textbf{Comparison of heat engine efficiencies in the coherent and incoherent modes of the quantum SWITCH for initially uncorrelated working medium and controller at $\beta\varepsilon = 0.1, 1, 10$.} \textit{Top row:} Efficiency gain $\Delta\eta_{\rm unc}^{\rm coh} = \tilde{\eta}_{\rm unc}^{\rm coh} - \eta_{\rm unc}^{\rm inc}$ as a function of the measurement strengths of $\mathbf{A}$ and $\mathbf{A}'$, where $\tilde{\eta}_{\rm unc}^{\rm coh}$ is optimized over the control state $\omega$, and $\eta_{\rm unc}^{\rm inc}$ is evaluated for the corresponding optimal $\omega$. \textit{Middle row:} Efficiency $\eta_{\rm unc}^{(-)}$ associated with the outcome $\ket{-}_{\rm c}$, optimized over $\omega$. \textit{Bottom row:} Efficiency $\eta_{\rm unc}^{(+)}$ for the outcome $\ket{+}_{\rm c}$, computed for the $\omega$ that maximizes $\eta_{\rm unc}^{(-)}$. In all panels, efficiencies are set to zero whenever the heat engine conditions on the measurement strengths are not satisfied.}
    \label{plt:opt_eta_unc}
\end{figure*}

We analyze how initial correlations between the working medium $\mathbf{Q}$ and the control $\mathbf{C}$ affect the performance of the heat engine implemented by the quantum SWITCH.

As a reference, we first recall the standard case studied in Ref.~\cite{dieguez2023thermal}, where $\mathbf{Q}$ and $\mathbf{C}$ are initially uncorrelated,
\begin{equation} \label{eq:rho^inc_unc}
    \varrho_{\rm in}^{\rm unc} = \rho^{(0)} \otimes \omega,
\end{equation}
where
\begin{equation}
\omega = \zeta \ketbrac{\psi_{\theta,\phi}} + (1-\zeta) \ketbrac{\psi_{\theta-\pi,\phi}},
\end{equation}
is a state of $\mathbf{C}$, with
\begin{equation}\label{app:pureState}
    \ket{\psi_{\theta,\phi}}_{\rm c} = \cos\left(\frac{\theta}{2}\right)\ket{0}_{\rm c}+\sin\left(\frac{\theta}{2}\right)e^{i\phi}\ket{1}_{\rm c}.
\end{equation}
Here, $\theta \in [0, \pi]$ and $\phi \in [0, 2\pi]$ parameterize the control's Bloch vector, while $\zeta \in [0,1]$ encodes the purity of $\omega$. In this setting, both the incoherent and coherent modes of the SWITCH act as generalized measurements whose outputs are diagonal in the energy basis of $\mathbf{Q}$ (see Appendix~\ref{app:uncorrelated}). Consequently, in both modes the second stroke can extract work only by permuting energy populations.

The incoherent mode implements an effective measurement strength
\begin{eqnarray}
    \nonumber \bar{a}^{\rm inc}_{\rm unc}(\theta) &=& \cos^2\Bigl(\frac{\theta}{2}\Bigr)a' + \sin^2\Bigl(\frac{\theta}{2}\Bigr)a,
\end{eqnarray}
with the corresponding population imbalance $\delta^{\rm inc}_{\rm unc}(\theta) = 1 - 2\bar{a}^{\rm inc}_{\rm unc}(\theta)$. In the coherent mode, SCO modifies the population imbalance from $\delta^{\rm inc}_{\rm unc}$ to $\delta^{\rm inc}_{\rm unc} \mp \tilde{\delta}_{\rm unc}^{\rm coh}$, yielding the maximal variation
\begin{align}
    \tilde{\delta}_{\rm unc}^{\rm coh}(\theta) &= \frac{1}{2}\Bigl((1-a)(1-a')(1-\tanh(\beta\varepsilon)) \nonumber \\
    &\qquad - aa'(1+\tanh(\beta\varepsilon))\Bigr)\sin(\theta),
\end{align}
which represents the largest SCO-induced change attainable for fixed $(a,a',\beta\varepsilon)$ (see Appendix~\ref{app:uncorrelated}). This allows us to express the non-negative SCO contribution to the efficiency as
\begin{equation}
    \Delta\eta^{\rm coh}_{\rm unc} = \frac{|\tilde{\delta}_{\rm unc}^{\rm coh}(\theta)| - |\delta_{\rm unc}^{\rm inc}(\theta)|}{\delta_{\rm unc}^{\rm inc}(\theta) + \tanh(\beta\varepsilon)}.
\end{equation}
To assess this contribution, we further optimize the efficiency $\tilde{\eta}_{\rm unc}^{\rm coh}$ over the control state $\omega$ (i.e., over $\theta$), thereby identifying the maximal performance of the heat engine in the absence of initial correlations between $\mathbf{Q}$ and $\mathbf{C}$ and comparing it with the performance of the incoherent mode under the same $\omega$ (see Fig.~\ref{plt:opt_eta_unc}).

\begin{figure*}[t!]
    \centering
    \includegraphics[width=\textwidth]{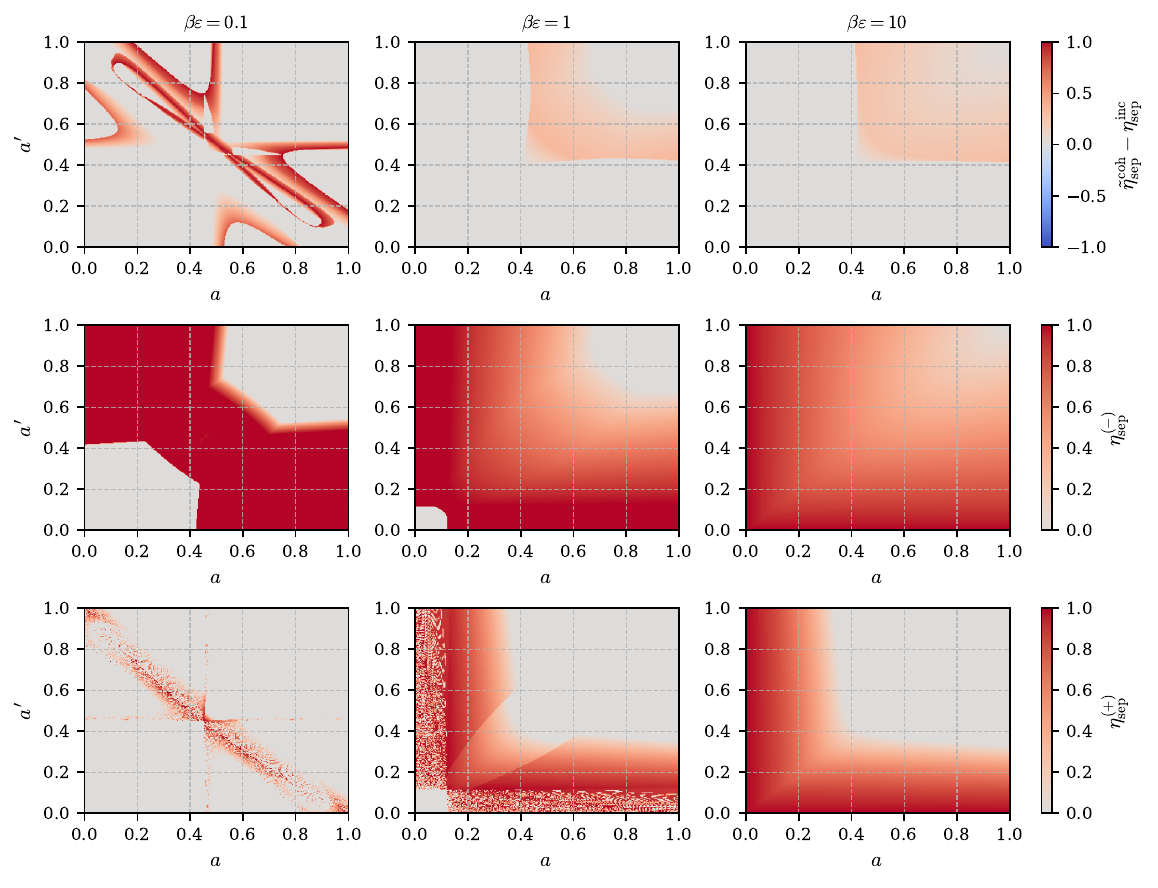}
    \caption{\textbf{Comparison of heat engine efficiencies in the coherent and incoherent modes of the quantum SWITCH for initially correlated separable states of the working medium and control system at $\beta\varepsilon = 0.1, 1, 10$.} \textit{Top row:} Efficiency gain $\Delta\eta_{\rm sep}^{\rm coh} = \tilde{\eta}_{\rm sep}^{\rm coh} - \eta_{\rm sep}^{\rm inc}$ as a function of the measurement strengths of $\mathbf{A}$ and $\mathbf{A}'$, where $\tilde{\eta}_{\rm sep}^{\rm coh}$ is optimized over the control state $\omega$, and $\eta_{\rm sep}^{\rm inc}$ is evaluated for the corresponding optimal $\omega$. \textit{Middle row:} Efficiency $\eta_{\rm sep}^{(-)}$ corresponding to the outcome $\ket{-}_{\rm c}$, optimized over $\omega$. \textit{Bottom row:} Efficiency $\eta_{\rm sep}^{(+)}$ corresponding to the outcome $\ket{+}_{\rm c}$, computed for the initial state that maximizes $\eta_{\rm sep}^{(-)}$. In all panels, efficiencies are set to zero whenever the heat engine conditions on the measurement strengths are not satisfied.}
    \label{plt:opt_eta_sep}
\end{figure*}

At high initial temperatures (small $\beta\varepsilon$), the range of admissible effective measurement strengths in the coherent mode is strongly constrained by the heat-engine conditions \eqref{eq:heatEngCondSwitch1}--\eqref{eq:heatEngCondSwitch2}. Nonetheless, notable efficiency gains appear when the measurement strengths $a$ and $a'$ are close to $\frac{1}{2}$ or lie along the symmetry line $a' = 1-a$, where interference between the causal orders is most pronounced. As the temperature decreases (larger $\beta\varepsilon$), the magnitude of the advantage over the incoherent mode diminishes, while the region supporting a positive gain broadens toward higher measurement strengths ($a, a' \gtrsim \frac{1}{2}$).

In a more general setting, the working medium $\mathbf{Q}$ and the control $\mathbf{C}$ may share an initially correlated state,
\begin{align}\label{eq:correlatedState}
    \varrho_{\rm in}^{\rm qe}
    &=
    \sum_{\ell = 0,1}
    \frac{1 + (-1)^\ell \tanh(\beta\varepsilon)}{2}\,
    \ketbra{\ell} \otimes \omega_{\ell} \nonumber \\
    &\qquad + \xi_{\beta\varepsilon}\, e^{-i\varphi}
    \ket{0}\bra{1} \otimes
    \ket{\psi_{\theta,\phi}}_{\rm c}\bra{\psi_{\theta-\pi,\phi}}
    + \text{h.c.},
\end{align}
which fulfills the local thermality condition~\eqref{eq:localThermalState}. Here, $\varphi \in [0, 2\pi]$, and
\begin{equation} \label{eq:omega}
    \omega_\ell =
    \zeta_\ell\, \ketbrac{\psi_{\theta,\phi}}
    + (1-\zeta_\ell)\, \ketbrac{\psi_{\theta-\pi,\phi}}
\end{equation}
represent the local contributions of $\mathbf{C}$ to the joint state. The parameters $\zeta_{0,1} \in [0, 1]$ and $\xi_{\beta\varepsilon} \in \Bigl[0, \frac{\operatorname{sech}(\beta\varepsilon)}{2}\Bigr]$ characterize the degree of correlations between $\mathbf{Q}$ and $\mathbf{C}$, which we explore in what follows. It is useful to distinguish, within this model, the roles of local purity and nonlocal correlations. In the incoherent mode, the dynamics of $\mathbf{Q}$ are fully determined by its reduced state: for fixed temperature and fixed energy populations, the heat and work exchanges depend only on the local purity of $\mathbf{Q}$. In this mode, correlations in $\varrho_{\rm in}$ do not influence the thermodynamic quantities beyond their effect on the marginals. In the coherent mode, by contrast, the conditional states \eqref{eq:firstStrokeStateSwitch} generated by the quantum SWITCH depend explicitly on the off-diagonal blocks of $\varrho_{\rm in}$ (see Appendix~\ref{app:Switch}). When correlations are present, their interplay with the SCO allows phase information stored in the joint state to be converted into coherence in $\mathbf{Q}$, thereby modifying the extractable work and the range of measurement strengths for which the engine operates.

\subsubsection{Initial correlations that allow a separable representation} \label{sub:ini-correlations}

We first consider the case where the working medium and the control system exhibit initial correlations but remain unentangled. In this scenario, $\xi_{\beta\varepsilon} = 0$, and the initial joint state is a convex mixture of uncorrelated product states. Without loss of generality, we impose $0 \leq \zeta_{1} \leq \zeta_{0} \leq 1$ and $\zeta_{0} \geq \frac{1}{2}$, so that $\zeta_{0} - \zeta_{1}$ quantifies the strength of separable correlations. Setting $\zeta_{0} - \zeta_{1} = 0$ reproduces the uncorrelated case discussed in Appendix~\ref{app:uncorrelated}, whereas $\zeta_{0} - \zeta_{1} = 1$ corresponds to maximally (yet still separably) correlated states.

\begin{figure*}[t!]
    \centering
    \includegraphics[width=\textwidth]{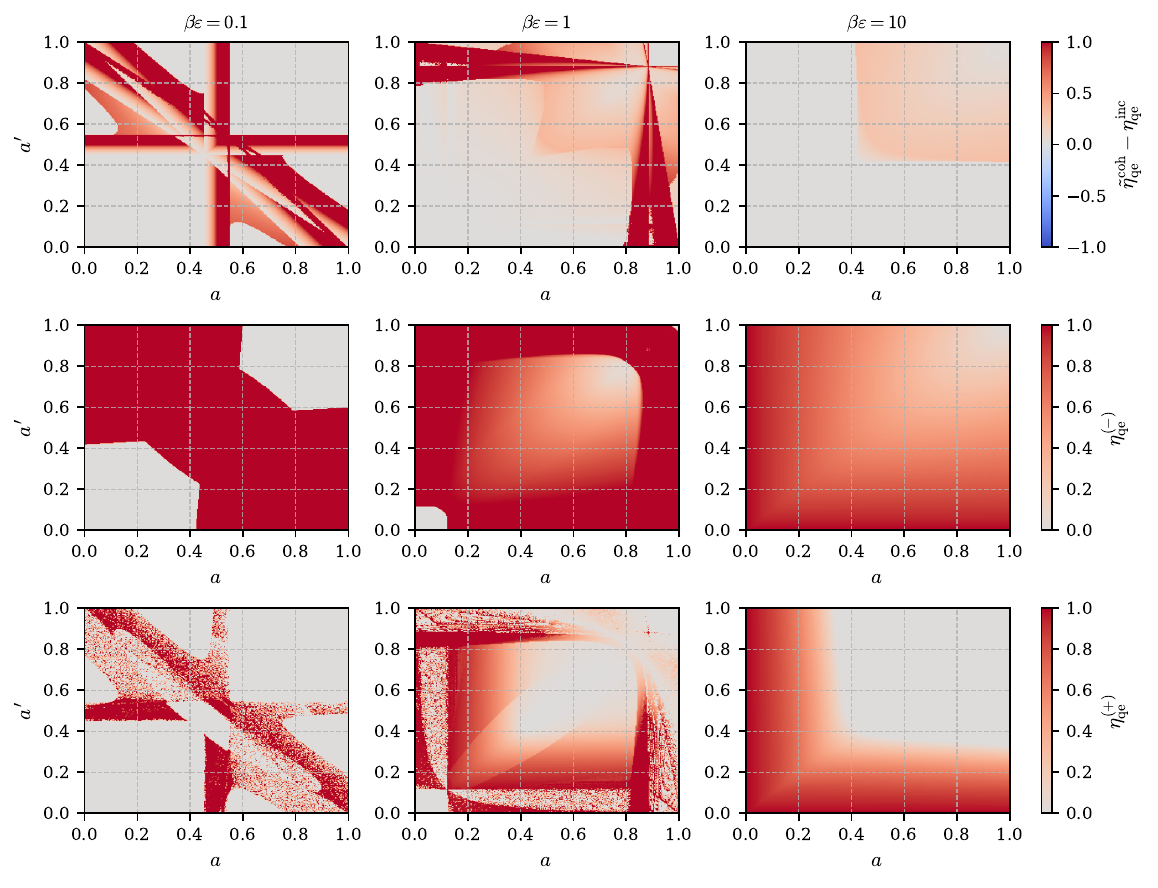}
    \caption{\textbf{Comparison of heat engine efficiencies in the coherent and incoherent modes of the quantum SWITCH for initially entangled working medium and controller at $\beta\varepsilon = 0.1, 1, 10$.} \textit{Top row:} Efficiency gain $\Delta\eta_{\rm qe}^{\rm coh} = \tilde{\eta}_{\rm qe}^{\rm coh} - \eta_{\rm qe}^{\rm inc}$ as a function of the measurement strengths of $\mathbf{A}$ and $\mathbf{A}'$, where $\tilde{\eta}_{\rm qe}^{\rm coh}$ is optimized over the control state $\omega$, and $\eta_{\rm qe}^{\rm inc}$ is evaluated for the corresponding optimal $\omega$. \textit{Middle row:} Efficiency $\eta_{\rm qe}^{(-)}$ associated with the outcome $\ket{-}_{\rm c}$, optimized over $\omega$. \textit{Bottom row:} Efficiency $\eta_{\rm qe}^{(+)}$ for the outcome $\ket{+}_{\rm c}$, computed for the initial state that maximizes $\eta_{\rm qe}^{(-)}$. In all panels, efficiencies are set to zero whenever the heat engine conditions on the measurement strengths are not satisfied.}
    \label{plt:opt_eta_ent}
\end{figure*}

Tracing out the control qubit yields a measurement channel acting on $\mathbf{Q}$ with an effective measurement strength
\begin{equation} \label{eq:a^inc_sep}
        \bar{a}^{\rm inc}_{\rm sep}(\theta, \zeta_{0,1}) = \bar{a}^{\rm inc}_{\rm unc}(\theta) + (a - a') \operatorname{tr} [\mathfrak{Z} \rho^{(0)}] \cos(\theta),
\end{equation}
with the corresponding population imbalance
\begin{equation}
    \delta^{\rm inc}_{\rm sep}(\theta, \zeta_{0,1}) = \delta^{\rm inc}_{\rm unc}(\theta) - 2(a - a') \operatorname{tr}[\mathfrak{Z} \rho^{(0)}] \cos(\theta).
\end{equation}
where $\mathfrak{Z} := \mathrm{diag}(\zeta_0,\zeta_1)$ and $\operatorname{tr}[\mathfrak{Z}\rho^{(0)}]$ quantifies the purity of the local state of $\mathbf{C}$, introducing dependence on the initial temperature of the working medium into the incoherent mode. A value of $\frac{1}{2}$ corresponds to a maximally mixed state, while $\operatorname{tr}[\mathfrak{Z}\rho^{(0)}] \in \{0,1\}$ identifies a pure control state (see Appendix~\ref{app:cc_purity}).

In the coherent mode, SCO alters the population imbalance according to
\begin{align}\label{app:a_bar_cl}
    \tilde{\delta}_{\rm sep}^{\rm coh}(\theta, \zeta_{0,1}) &= (\zeta_0 + \zeta_1 - 1)\delta_{\rm unc}^{\rm coh}(\theta) \nonumber \\
    & - \frac{\zeta_0 - \zeta_1}{2}\Bigl[(1-a)(1-a')(1-\tanh(\beta\varepsilon)) \nonumber \\
    &\qquad + aa'(1+\tanh(\beta\varepsilon))\Bigr]\sin(\theta),
\end{align}
representing the maximal SCO-induced modification attainable for fixed $(a, a', \beta\varepsilon)$ after optimization over the control phase $\phi$. The corresponding non-negative SCO contribution to the efficiency reads
\begin{equation}
    \Delta\eta^{\rm coh}_{\rm sep} = \frac{|\tilde{\delta}_{\rm sep}^{\rm coh}(\theta, \zeta_{0,1})| - |\delta_{\rm sep}^{\rm inc}(\theta, \zeta_{0,1})|}{\delta_{\rm sep}^{\rm inc}(\theta, \zeta_{0,1}) + \tanh(\beta\varepsilon)}.
\end{equation}
Optimization of $\tilde{\eta}_{\rm sep}^{\rm coh}$ over $\zeta_{0,1}$ and $\theta$ reveals two distinct configurations yielding maximal performance: (i) perfectly separable correlations, $\zeta_0 - \zeta_1 = 1$, and (ii) a pure but uncorrelated control state, $\zeta_0 = \zeta_1 = 1$ (see Appendix~\ref{app:cc_purity}). Comparison with the incoherent mode under an identical initial state (see Fig.~\ref{plt:opt_eta_sep}) shows that separable correlations enhance efficiency primarily along the symmetry line $a' = 1 - a$ at high initial temperatures (small $\beta\varepsilon$). As the temperature decreases (larger $\beta\varepsilon$), the advantage localizes near $a \simeq a' \simeq \frac{1}{2}$ and gradually converges to the performance of the uncorrelated case in the low-temperature limit. Hence, separable correlations can assist SCO in enhancing engine performance at high temperatures by expanding the parameter range $(a,a')$ that yields a positive efficiency gain over operations with definite or probabilistic causal order.

\begin{figure}[t!]
    \centering
    \includegraphics[width=\columnwidth]{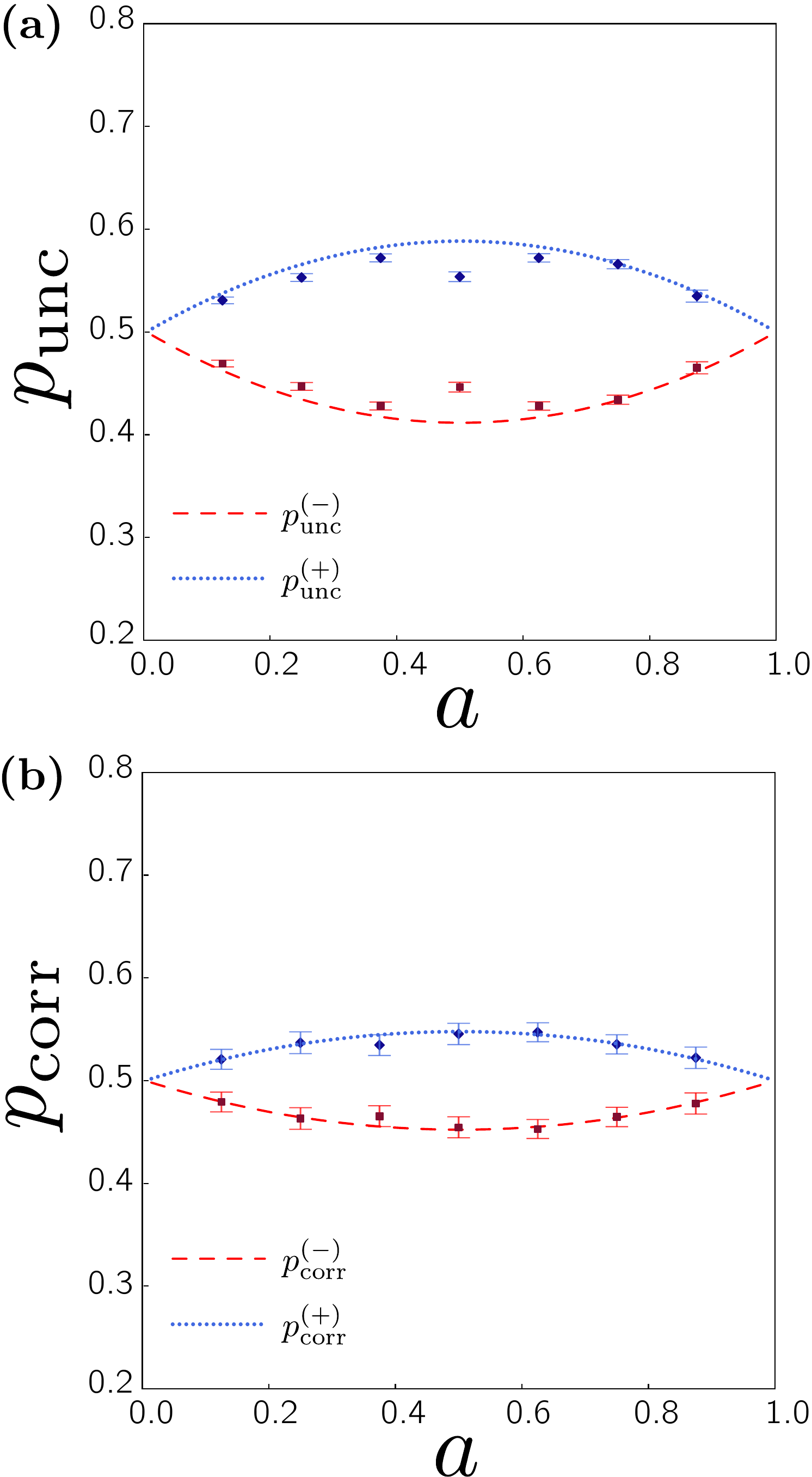}
    \caption{\textbf{Comparison of measurement outcome probabilities for different initial states.} Measurement outcome probabilities $p^{(\pm)}$ of the control qubit $\mathbf{C}$ are shown as functions of the measurement strength $a$. The initial states satisfy $\theta = \frac{\pi}{2}$, $\phi = \frac{\pi}{4}$, $\phi' = 0$, and $(\beta\varepsilon)^{-1} = \Texp$, and the measurement strengths satisfy $a' = 1 - a$. Experimental data are shown as symbols with error bars indicating statistical uncertainty, and the markers denote mean values. The red and blue elements correspond to the probabilities of the outcomes $\ket{-}_{\rm c}$ and $\ket{+}_{\rm c}$, respectively, shown for \textbf{(a)} the uncorrelated initial state and \textbf{(b)} the entangled initial state. Notice that the probabilities $p^{(\pm)}$ do not depend on the contribution proportional to $\xi_{\beta\varepsilon}$ in \eqref{eq:correlatedState}. Therefore, separable and entangled initial states lead to identical statistics for $p^{(\pm)}$, and only the data for the entangled case are displayed in panel~(b).}
    \label{plt:prob}
\end{figure}

\subsubsection{Initial entanglement}

In the more general case $\xi_{\beta\varepsilon} \geq 0$, the working medium $\mathbf{Q}$ and the controller $\mathbf{C}$ can be initially entangled. For perfect correlations ($\zeta_0 - \zeta_1 = 1$), the initial state~\eqref{eq:correlatedState} continuously interpolates between a separable configuration at $\xi_{\beta\varepsilon} = 0$ and a pure entangled state
\begin{equation}
    \varrho_{\rm in}^{\mathrm{qe}} = \ketbra{\Psi_{\theta,\phi}(\beta\varepsilon)},
\end{equation}
with
\begin{eqnarray}
        \ket{\Psi_{\theta,\phi}(\beta\varepsilon)} &=& \sqrt{\frac{1 +  \tanh(\beta\varepsilon)}{2}} \ket{0} \otimes \ket{\psi_{\theta, \phi}}_{\rm c} \nonumber \\
        &+& e^{-i\varphi}\sqrt{\frac{1 -  \tanh(\beta\varepsilon)}{2}} \ket{1} \otimes \ket{\psi_{\theta-\pi, \phi}}_{\rm c},
\end{eqnarray}
achieved for $\xi_{\beta\varepsilon} = \frac{\operatorname{sech}(\beta\varepsilon)}{2}$.

In the incoherent mode, tracing out the control qubit $\mathbf{C}$ yields the same effective measurement strength as in the corresponding separable case obtained by setting $\xi_{\beta\varepsilon} = 0$,
\begin{equation}
    \bar{a}_{\rm qe}^{\rm inc}(\theta, \zeta_{0,1}, \xi_{\beta\varepsilon})
    = \bar{a}_{\rm sep}^{\rm inc}(\theta, \zeta_{0,1}).
\end{equation}
In contrast, in the coherent mode the SCO not only modifies the population imbalance in the working medium $\mathbf{Q}$ but also generates quantum coherence in its state (see Appendix~\ref{app:cc_purity}). This coherence enhances the extractable work and enlarges the parameter region $(a,a')$ in which the engine operates, becoming particularly relevant at high temperatures (small $\beta\varepsilon$, see Fig.~\ref{plt:opt_eta_ent}). The corresponding optimal efficiency
\begin{equation}\label{eq:effEntCoh}
    \Delta\eta_{\rm qe}^{\rm coh} = \frac{\sqrt{(\delta_{\rm sep}^{\rm inc}(\theta, \zeta_{0,1}))^2 + |\tilde{\xi}(\theta, \xi_{\beta\varepsilon})|^2} + \delta_{\rm sep}^{\rm inc}(\theta, \zeta_{0,1})}{\delta_{\rm sep}^{\rm inc}(\theta, \zeta_{0,1}) + \tanh(\beta\varepsilon)},
\end{equation}
is governed by the state coherence, which contributes via
\begin{equation}
\tilde{\xi}(\theta, \xi_{\beta\varepsilon}) = 2\xi_{\beta\varepsilon}\Bigl(a(1-a')\sin^2\Bigl(\frac{\theta}{2}\Bigr) + a'(1-a)\cos^2\Bigl(\frac{\theta}{2}\Bigr)\Bigr).
\end{equation}

\begin{figure*}[t!]
    \centering
    \includegraphics[width=\linewidth]{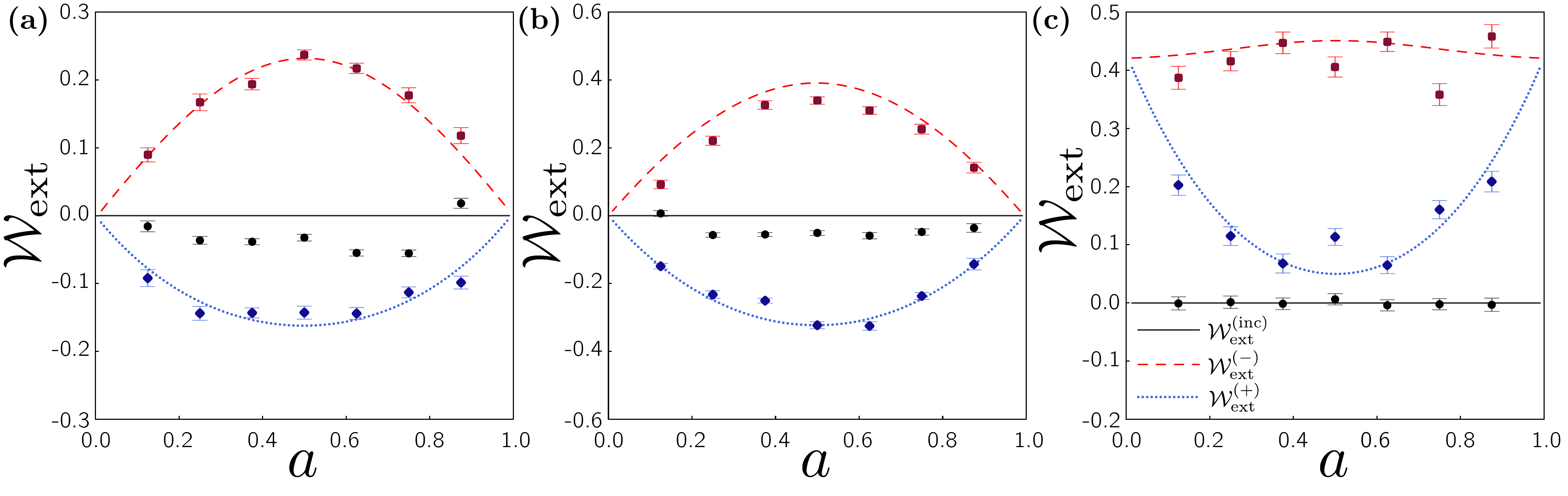}
    \caption{\textbf{Extracted work conditioned on the measurement outcome of the controller.} Extracted work $\mathcal{W}_{\rm ext}^{(\pm)}$ as a function of the measurement strength $a$. The initial parameters are fixed to $\theta = \frac{\pi}{2}$, $\phi = \frac{\pi}{4}$, $\phi' = 0$, and $(\beta\varepsilon)^{-1} = \Texp$, and the measurement strengths satisfy $a' = 1 - a$. Experimental data are shown as symbols with error bars indicating statistical uncertainty, where markers denote mean values. The red and blue elements correspond to $\mathcal{W}_{\rm ext}^{(-)}$ and $\mathcal{W}_{\rm ext}^{(+)}$, respectively, while black elements represent the work extracted in the incoherent mode of the quantum SWITCH. Panels correspond to the joint initial state of $\mathbf{Q}$ and $\mathbf{C}$ being \textbf{(a)} uncorrelated, \textbf{(b)} separable, and \textbf{(c)} entangled.}
    \label{plt:work}
\end{figure*}

Decreasing the initial temperature of $\mathbf{Q}$ (increasing $\beta\varepsilon$) shrinks the $(a,a')$ region where the coherence-dominated efficiency~\eqref{eq:effEntCoh} is optimal, pushing it toward extreme measurement strengths. Outside this domain, the performance gain arises from the interplay between population imbalance and coherence, yielding
\begin{widetext}
\begin{eqnarray}\label{eq:effInterplayCoh}
    \Delta\eta_{\rm qe}^{\rm coh} &=& \frac{\frac{1}{2}\sum_\pm \sqrt{(\delta_{\rm sep}^{\rm inc}(\theta, \zeta_{0,1}) \pm \tilde{\delta}_{\rm sep}^{\rm coh}(\theta, \zeta_{0,1}))^2 + |\tilde{\xi}'(\theta, \xi_{\beta\varepsilon})|^2} + \delta_{\rm sep}^{\rm inc}(\theta, \zeta_{0,1})}{\delta_{\rm sep}^{\rm inc}(\theta, \zeta_{0,1}) + \tanh(\beta\varepsilon)},
\end{eqnarray}
\end{widetext}
where the coherence contribution enters as
\begin{equation}
\tilde{\xi}'(\theta, \xi_{\beta\varepsilon}) = 2\xi_{\beta\varepsilon}\Bigl(a(1-a')\sin^2\Bigl(\frac{\theta}{2}\Bigr) - a'(1-a)\cos^2\Bigl(\frac{\theta}{2}\Bigr)\Bigr).
\end{equation}
Finally, we observe that, similarly to the case of separable initial states, the optimal efficiency converges toward that of the uncorrelated configuration in the low-temperature limit.

In both regimes \eqref{eq:effEntCoh} and \eqref{eq:effInterplayCoh}, the efficiency increases monotonically with the correlation strength $\xi_{\beta\varepsilon}$, provided that the heat engine constraints \eqref{eq:heatEngCondSwitch1}--\eqref{eq:heatEngCondSwitch3} remain satisfied. This indicates that prior entanglement between the working medium and the control not only enlarges the range of measurement strengths $(a,a')$ enabling an efficiency advantage over the incoherent mode but also enhances the achievable efficiency beyond what is attainable with uncorrelated or merely separable initial states.


\subsection{Proof-of-principle experiment}\label{subsec:exp}

Following our theoretical analysis of the heat engine's performance, we present its simulation on the IBM Quantum Experience platform as a proof-of-principle demonstration (see Appendix~\ref{sec:metods_error} for methodological details). To investigate the role of initial correlations between the working medium $\mathbf{Q}$ and the control system $\mathbf{C}$, we consider idealized preparations of the three representative initial configurations introduced in Section~\ref{sub:ini-correlations}:
\begin{enumerate}[(i)]
    \item an uncorrelated product state $\varrho_{\rm in}^{\rm unc}$ with a pure control state ($\zeta = 1$),
    \item a separable state $\varrho_{\rm in}^{\rm sep}$ with perfect correlations ($\zeta_0 = 1$, $\zeta_1 = 0$), and
    \item an entangled state $\varrho_{\rm in}^{\mathrm{qe}}$ with perfect correlations ($\zeta_0 = 1$, $\zeta_1 = 0$, $\xi_{\beta\varepsilon} = \frac{\operatorname{sech}(\beta\varepsilon)}{2}$).
\end{enumerate}

To isolate the contribution of the SCO, we choose the parameters such that no work can be extracted in the incoherent regime, i.e., $\eta^{\rm inc} = 0$. This condition is achieved by setting $\theta = \frac{\pi}{2}$ and employing complementary measurement apparata $\mathbf{A}$ and $\mathbf{A}'$ with $a' = 1 - a$, ensuring $\bar{a}^{\rm inc} = \frac{1}{2}$ for all three initial configurations. In addition, the simulation is performed with $\phi = \frac{\pi}{4}$ and a temperature fulfilling $(\beta\varepsilon)^{-1} = \Texp$.

We begin by examining the extractable work $\mathcal{W}_{\rm ext}^{(\pm)}$ from $\mathbf{Q}$ in the second stroke, conditioned on the measurement outcome $\ket{\pm}_{\rm c}$ of $\mathbf{C}$. The corresponding outcome probabilities (see Appendix~\ref{app:Switch}) are shown in Fig.~\ref{plt:prob}. Notably, the separable and entangled initial states produce identical statistics, whereas the uncorrelated configuration exhibits a stronger outcome bias,
\begin{equation}
    |2p^{(\pm)}_{\rm sep/qe} - 1| < |2p^{(\pm)}_{\rm unc} - 1|,
\end{equation}
indicating that correlations reduce the asymmetry between the two outcomes.

\begin{figure}[t!]
    \centering
    \includegraphics[width=\columnwidth]{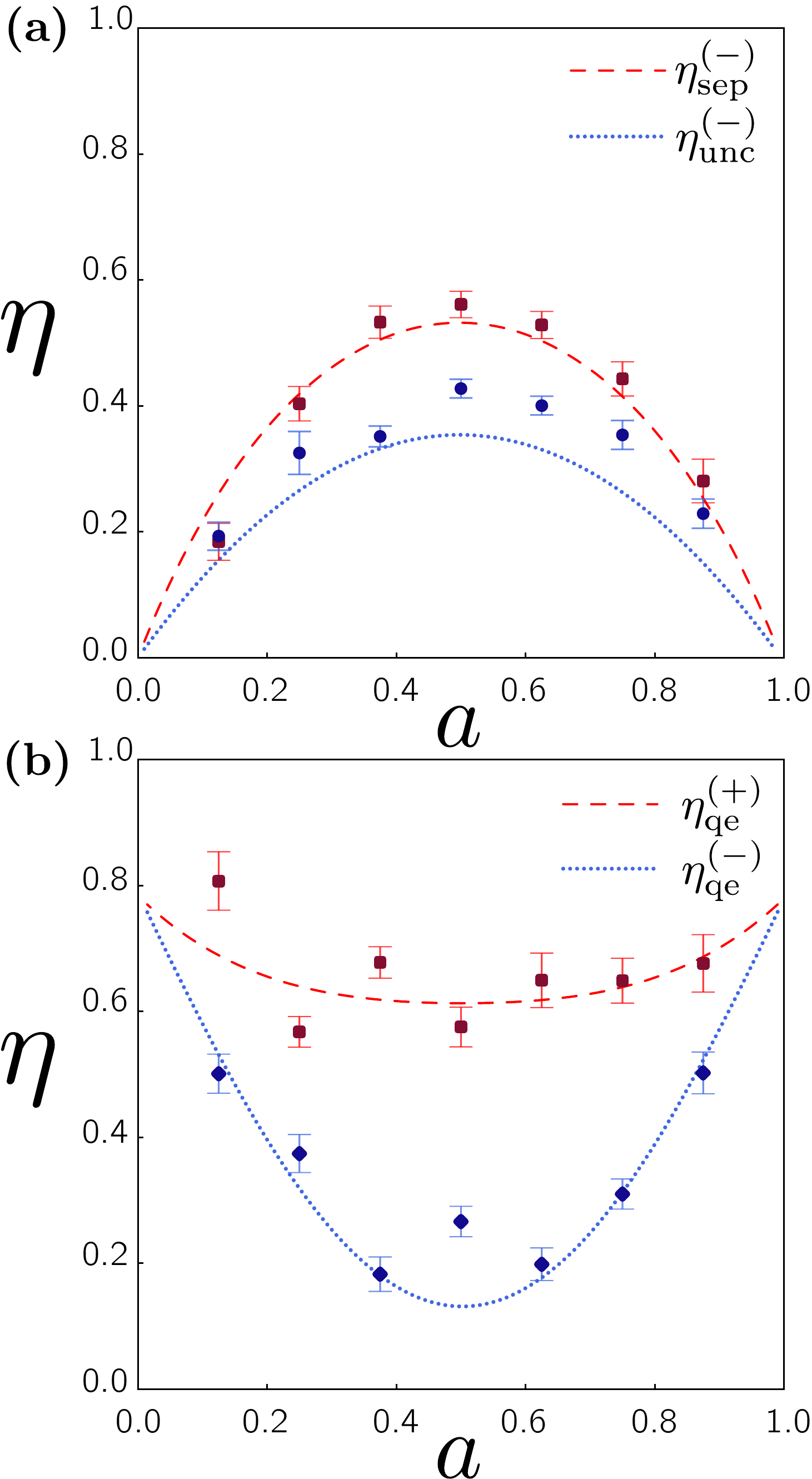}
    \caption{\textbf{Efficiencies conditioned on the measurement outcome of the controller.} Efficiencies $\eta^{(\pm)}$ as functions of the measurement strength $a$. The initial parameters are fixed to $\theta = \frac{\pi}{2}$, $\phi = \frac{\pi}{4}$, $\phi' = 0$, and $(\beta\varepsilon)^{-1} = \Texp$, and the measurement strengths satisfy $a' = 1 - a$. Experimental data are shown as symbols with error bars indicating statistical uncertainty, and markers denote mean values. \textbf{(a)} Red and blue elements correspond to the efficiency $\eta^{(-)}$ associated with the outcome $\ket{-}_{\rm c}$ for separable and uncorrelated initial states, respectively. \textbf{(b)} Red and blue elements correspond to the efficiencies $\eta^{(+)}$ and $\eta^{(-)}$ associated with the outcomes $\ket{+}_{\rm c}$ and $\ket{-}_{\rm c}$, respectively, for an entangled initial state.}
    \label{plt:eta}
\end{figure}

In our setting, for uncorrelated and separable initial states, work can be extracted only when the measurement of $\mathbf{C}$ yields the outcome $\ket{-}_{\rm c}$. The opposite outcome violates the heat engine conditions~\eqref{eq:heatEngCondSwitch1}--\eqref{eq:heatEngCondSwitch3}, leading to work being invested rather than extracted. Nevertheless, assistance of the SCO by separable correlations enhances the work extracted under the outcome $\ket{-}_{\rm c}$ compared with the uncorrelated case, attaining its maximum at $a = a' = \frac{1}{2}$. In this regime, the SCO contribution vanishes at the boundary points $a \in \{0,1\}$, coinciding with the incoherent mode. By contrast, initial entanglement between $\mathbf{Q}$ and $\mathbf{C}$ enables work extraction even at the boundaries by consuming the coherence generated through the quantum SWITCH, allowing both measurement outcomes of the controller to yield positive extracted work and enhancing work extraction beyond that attainable in the uncorrelated and separable cases (see Fig.~\ref{plt:work}).

To assess the heat engine's performance, we reconstruct its efficiency via state tomography. We first analyze the efficiencies~\eqref{eq:effPS} associated with the individual measurement outcomes of $\mathbf{C}$, which are relevant whenever the feedback cost can be neglected. As shown in Fig.~\ref{plt:eta}, these efficiencies exhibit the hierarchy $\eta^{(+)}_{\rm qe} > \eta^{(-)}_{\rm sep} > \eta^{(-)}_{\rm unc}$ across the full range of measurement strengths. Comparison between the entangled and the other regimes reveals that, in the former, the outcome $\ket{+}_{\rm c}$ contributes positively to engine operation, while the efficiency profile inverts (attaining its minimum at $a = a' = \frac{1}{2}$), thus emphasizing the role of coherence in the interplay between SCO and initial correlations.

\begin{figure}[t!]
    \centering
    \includegraphics[width=\columnwidth]{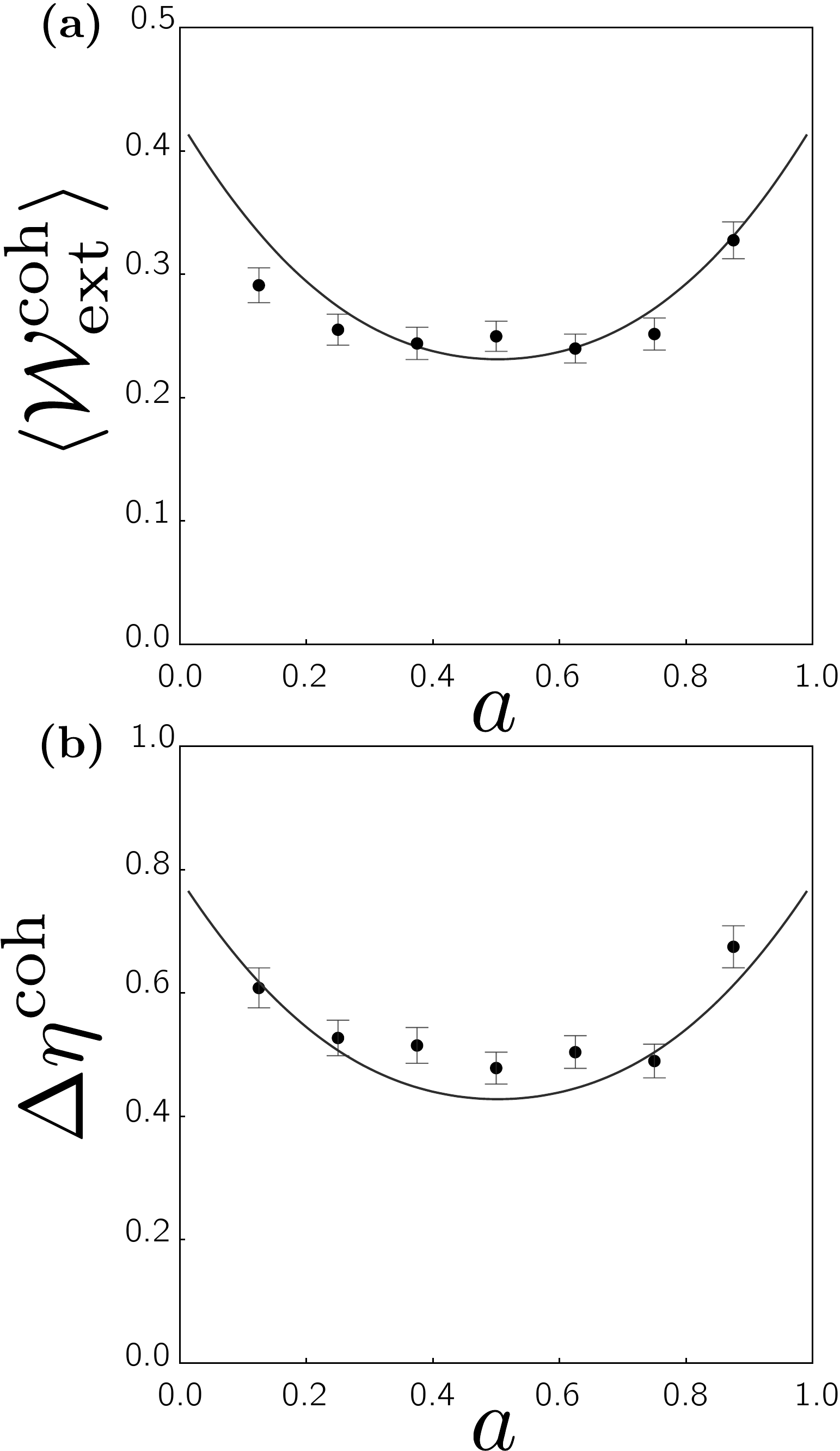}
    \caption{\textbf{Average extracted work and SCO contribution to the efficiency under an initial entangled state.} \textbf{(a)} Extracted work $\langle \mathcal{W}_{\rm ext}^{\rm coh} \rangle$ and \textbf{(b)} SCO contribution $\Delta\eta^{\rm coh}$ to the heat engine efficiency as functions of the measurement strength $a$. The initial parameters are fixed to $\theta = \frac{\pi}{2}$, $\phi = \frac{\pi}{4}$, $\phi' = 0$, and $(\beta\varepsilon)^{-1} = \Texp$, and the measurement strengths satisfy $a' = 1 - a$. Experimental data are shown as symbols with error bars indicating statistical uncertainty, and markers denote mean values.}
    \label{plt:mean}
\end{figure}

Since positive work extraction occurs for both measurement outcomes in the entangled case, we can consider the average extracted work~\eqref{eq:extWorkSwitch} and the corresponding efficiency~\eqref{eq:etaSwitch}, which simplifies to
\begin{equation}
    \eta^{\rm coh} = \Delta\eta^{\rm coh} - \eta_{\rm cost},
\end{equation}
given that the incoherent mode has zero efficiency in our setting. Consequently, the heat engine achieves a non-zero efficiency whenever
\begin{equation}
    \langle\mathcal{W}_{\rm ext}^{\rm coh} \rangle > - \mathcal{W}_{\rm cost},
\end{equation}
where $\mathcal{W}_{\rm cost}$ denotes the minimal erasure cost, defined in~\eqref{eq:workCost} according to Landauer's principle. The average work $\langle\mathcal{W}_{\rm ext}^{\rm coh} \rangle$ and the SCO-induced efficiency $\Delta\eta^{\rm coh}$ are displayed in Fig.~\ref{plt:mean}. Since $\mathcal{W}_{\rm cost}$ depends solely on the detector's temperature $T_{\rm D}$, it remains constant across all measurement strengths.

\section{Outlook and conclusions} \label{sec:conclusions}

We have investigated a measurement-driven quantum heat engine in which energy exchanges are governed by measurements applied in a superposition of causal orders. Our analysis examined how work extraction and overall engine performance are affected when the working medium and the controller share prior correlations. Three distinct regimes were considered: (i) uncorrelated initial states, (ii) classically correlated separable states, and (iii) entangled states.

The results show that the engine's performance systematically improves with the strength of initial correlations. While separable correlations already enhance the amount of extractable work compared with the uncorrelated case, entanglement provides a qualitatively distinct advantage. When assisted by entanglement, the superposition of causal orders generates coherence in the working medium prior to the work-extraction stroke, thereby enabling operation in parameter regions where definite-order dynamics would yield zero efficiency.

To analyze the branch-dependent effects of the coherent control, we evaluated both the efficiencies conditioned on individual measurement outcomes and the overall efficiency defined from the average extracted work. The latter accounts for contributions from both control outcomes whenever each yields a thermodynamic advantage in the form of positive extractable work. These theoretical predictions were verified through a proof-of-principle simulation on the IBM Quantum Experience platform, where generalized measurement channels were implemented in a superposition of causal orders. The experimental results confirmed that coherently superposing the order of measurement channels activates work extraction and enhances efficiency beyond what is achievable under definite causal order.

These findings demonstrate not only the superior performance of cycles operating under superposed causal order compared with their definite-order counterparts, but also the operational significance of different forms of initial correlations. In particular, our results establish a concrete connection between initial correlations, coherence generation, and thermodynamic enhancement enabled by superposed causal order of operations. The present framework paves the way for exploring more complex thermodynamic setups (such as refrigeration cycles or feedback-controlled processes) where quantum correlations and indefinite causal order may jointly contribute to improved thermodynamic performance. Such extensions will help clarify the broader role of indefinite causal structures and their interplay with quantum resources in nonequilibrium thermodynamics.

\section*{Acknowledgments}
We thank André H. A. Malavazi for discussions. V.F.L. and R.M.S. acknowledge the support from CNPq, CAPES (grant no. 88887.712952/2022-00), and FAPESP.  P.R.D. acknowledges support from the NCN Poland, ChistEra-2023/05/Y/ST2/00005 under the project Modern Device Independent Cryptography (MoDIC). K.S. acknowledges: This research was funded in whole or in part by the Austrian Science Fund (FWF) 10.55776/PAT4559623. For open access purposes, the author has applied a CC BY public copyright license to any author-accepted manuscript version arising from this submission.

\appendix   


\begin{table*}[ht!]
\centering
\caption{Variations of internal energy and entropy during each stroke of the heat engine operating under a definite causal order of measurements.}
\label{tab:energy1}
\hspace*{-1cm}
\begin{tabular}{|l|ll|ll|}
    \hline
    &\multicolumn{2}{l}{Internal energy variation}&\multicolumn{2}{|l|}{Entropy variation} \\
    \hline
    \textbf{Stroke 1:} & $\Delta U^{(1)}$ &$= \varepsilon (1 - 2a + \tanh(\beta \varepsilon))$ & $\Delta S^{(1)}$ &$= \mathcal{H}_\text{bin}(a) - \mathcal{H}_\text{bin}\left(\frac{1 + \tanh(\beta \varepsilon)}{2}\right)$ \\
    \textbf{Stroke 2:} & $\Delta U^{(2)}$ &$= -\varepsilon (1 - 2a + |1-2a|)$ & $\Delta S^{(2)}$ &$= 0$ \\
    \textbf{Stroke 3:} & $\Delta U^{(3)}$ &$= \varepsilon (|1-2\bar{a}| - \tanh(\beta \varepsilon))$ & $\Delta S^{(3)}$ &$= \mathcal{H}_\text{bin}\left(\frac{1 + \tanh(\beta \varepsilon)}{2}\right) - \mathcal{H}_\text{bin}(a)$ \\
    \hline
\end{tabular}
\end{table*}

\section{Energy and entropy in the heat engine}
\label{app:heatEngVar}

Before the heat engine starts, the working medium $\mathbf{Q}$ is initialized in a Gibbs state
\begin{equation}\label{app:eq:GibbsState}
\rho^{(0)} = \frac{1}{2}\operatorname{diag}\Bigl[1+\tanh(\beta\varepsilon), \, 1-\tanh(\beta\varepsilon)\Bigr],
\end{equation}
at inverse temperature $\beta$, characterized by internal energy and entropy
\begin{eqnarray}
\label{app:eq:zeroEnergy} U^{(0)} &=& -\varepsilon \tanh(\beta\varepsilon), \\
\label{app:eq:zeroEntropy} S^{(0)} &=& \mathcal{H}_{\rm bin}\Bigl(\frac{1}{2}(1+\tanh(\beta\varepsilon))\Bigr),
\end{eqnarray}
where $\mathcal{H}_{\rm bin}(x)$ denotes the binary entropy function.

\emph{First stroke: heat injection.}
The first stroke consists of a non-selective measurement of $\mathbf{Q}$ performed by apparatus $\mathbf{A}$, implementing the channel $\mathcal{M}^a$. After this operation, the working medium is left in the state
\begin{eqnarray}
    \rho^{(1)} &=& \operatorname{diag}[a, 1-a].
\end{eqnarray}
with
\begin{eqnarray}
    U^{(1)} &=& \varepsilon (1-2a), \\
    S^{(1)} &=& \mathcal{H}_{\mathrm{bin}} (a).
\end{eqnarray}
Hence, the corresponding variations in internal energy and entropy are
\begin{eqnarray}
    \Delta U^{(1)} &=& \varepsilon \Bigl( 1 - 2a +  \tanh(\beta\varepsilon) \Bigr), \\
    \Delta S^{(1)} &=& \mathcal{H}_{\mathrm{bin}} (a) - \mathcal{H}_{\mathrm{bin}} \Bigl(\frac{1}{2}(1+\tanh(\beta\varepsilon))\Bigr),
\end{eqnarray}
To interpret this energy change as heat $\mathcal{Q}_{\rm hot}$ absorbed from $\mathbf{A}$, one requires $U^{(1)}\geq U^{(0)}$ and $S^{(1)} > S^{(0)}$, giving
\begin{eqnarray}
    \mathcal{Q}_{\rm hot} &=& \varepsilon \Bigl( 1 - 2a +  \tanh(\beta\varepsilon) \Bigr), \\
    \label{eq:app:aLowBound} a &>& \frac{1}{2}\Bigl( 1 - \tanh(\beta\varepsilon) \Bigr), \\
    \label{eq:app:aUppBound} a &<& \frac{1}{2}\Bigl( 1 + \tanh(\beta\varepsilon) \Bigr).
\end{eqnarray}

\emph{Second stroke: work extraction.}
In the second stroke, $\mathbf{Q}$ is measured by apparatus $\mathbf{B}$ via $\mathcal{M}^b$, yielding
\begin{eqnarray}
    \rho^{(2)} &=& \operatorname{diag}[b, 1-b],
\end{eqnarray}
with
\begin{eqnarray}
    U^{(2)} &=& \varepsilon (1-2b), \\
    S^{(2)} &=& \mathcal{H}_{\mathrm{bin}} (b).
\end{eqnarray}
This produces
\begin{eqnarray}
    \Delta U^{(2)} &=& \varepsilon ( a - b ), \\
    \Delta S^{(2)} &=& \mathcal{H}_{\mathrm{bin}} (b) - \mathcal{H}_{\mathrm{bin}} (a).
\end{eqnarray}
If the measurement is isentropic, $\Delta S^{(2)}=0$, then $|1-2b| = |1-2a|$, equivalently $|U^{(2)}| = |U^{(1)}|$, or more generally
\begin{equation}
    U^{(2)} = \Gamma^{(2)}|U^{(1)}|,
\end{equation}
where $\Gamma^{(2)} \in \{-1,1\}$ specifies the isentropic branch of the second stroke. To ensure that this energy change represents work extracted from $\mathbf{Q}$, the internal energy must decrease, i.e., $U^{(2)} \leq U^{(1)}$, implying that $\mathcal{W}_{\rm ext} = -\Delta U^{(2)}$, and
\begin{eqnarray}
    \nonumber \mathcal{W}_{\rm ext} &=& \varepsilon \bigl( U^{(1)} - \Gamma^{(2)}|U^{(1)}| \bigr) \\
    &=& \varepsilon \Bigl( 1-2a - \Gamma^{(2)}|1-2a| \Bigr), \\
    b &=& \frac{1}{2}\Bigl( 1 - \Gamma^{(2)}|1-2a| \Bigr).
\end{eqnarray}
The parameter $\Gamma^{(2)}$ specifies the operational branch of the second stroke: work is \textit{invested} ($\Gamma^{(2)} = 1$, $\mathcal{W}_{\rm ext} \leq 0$) or \textit{extracted} ($\Gamma^{(2)} = -1$, $\mathcal{W}_{\rm ext} \geq 0$) from $\mathbf{Q}$. Indeed, no work is produced for $\Gamma^{(2)} = 1$ and $U^{(2)} = U^{(1)}$. Hence, focusing on the work-extracting branch $\Gamma^{(2)} = -1$ yields
\begin{eqnarray}
    \mathcal{W}_{\rm ext} &=& \varepsilon \Bigl( 1-2a + |1-2a| \Bigr), \\
    b &=& \frac{1}{2}\Bigl( 1 + |1-2a| \Bigr).
\end{eqnarray}

\emph{Third stroke: thermalization.}
Finally, the medium $\mathbf{Q}$ is thermalized with the cold reservoir $\mathbf{R}$, completing the cycle with $\rho^{(3)}=\rho^{(0)}$. The energy and entropy variations in this stroke are
\begin{eqnarray}
    \Delta U^{(3)} &=& -\varepsilon \Bigl( \tanh(\beta\varepsilon) + 1 - 2b\Bigr), \\
    \Delta S^{(3)} &=& \mathcal{H}_{\mathrm{bin}} \Bigl(\frac{1}{2}(1+\tanh(\beta\varepsilon))\Bigr) - \mathcal{H}_{\mathrm{bin}} (b).
\end{eqnarray}
Similarly to the first stroke, the energy exchange is interpreted as heat $\mathcal{Q}_{\rm cold}$, yet transported from $\mathbf{Q}$ to $\mathbf{R}$. For this purpose, we set $\mathcal{Q}_{\rm cold} = \Delta U^{(3)} \leq 0$, which leads to
\begin{equation}
    \mathcal{Q}_{\rm cold} = \varepsilon \Bigl( |1-2a| - \tanh(\beta\varepsilon)\Bigr),
\end{equation}
under the constraints~\eqref{eq:app:aLowBound}--\eqref{eq:app:aUppBound} on the measurement strength $a$. The three strokes are summarized in Table~\ref{tab:energy1}.


\section{Extended heat engine with SCO}
\subsection{Operations in SCO: Quantum SWITCH}\label{app:Switch}

\begin{figure}[ht!]
    \centering
    \includegraphics[width=\columnwidth]{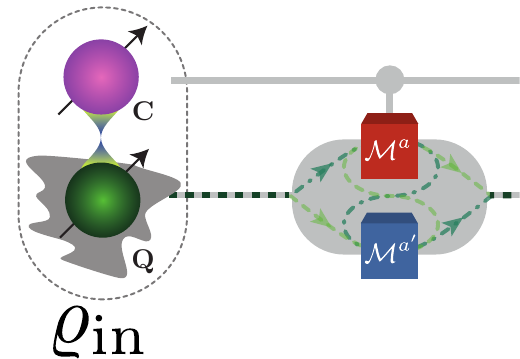}
    \caption{\textbf{Quantum SWITCH.} The setup consists of the working medium $\mathbf{Q}$ and the control qubit $\mathbf{C}$ prepared in an initial joint state $\varrho_{\rm in}$. The working medium undergoes two measurement channels, $\mathcal{M}^a[\cdot]$ and $\mathcal{M}^{a'}[\cdot]$, implemented by the apparata $\mathbf{A}$ and $\mathbf{A}'$, respectively. The control qubit determines their causal order: when $\mathbf{C}$ is in state $\ket{0}_{\rm c}$, $\mathbf{A}'$ acts after $\mathbf{A}$, while for $\ket{1}_{\rm c}$, the order is reversed. A coherent superposition of these states places the measurements in SCO.}
    \label{fig:switchSetup}
\end{figure}

The heat engine described in Section~\ref{sec:HeatEngine} consists of a sequence of strokes corresponding to quantum operations executed in a definite order. For instance, performing a measurement of the working medium $\mathbf{Q}$ first with apparatus $\mathbf{A}$ and subsequently with $\mathbf{A}'$ corresponds to the ordered composition of channels $\mathcal{M}^{a'} \circ \mathcal{M}^{a}$. However, quantum mechanics allows one to go beyond such ordered combinations of operations, e.g., placing the causal order itself in quantum superposition.

A paramount example is provided by the quantum SWITCH, a higher-order transformation that coherently controls the order of two operations, thus producing an SCO incompatible with any causally ordered combination of them. Given two quantum operations $\mathcal{A}[\cdot]$ and $\mathcal{B}[\cdot]$, the setup involves a target system that undergoes these operations and a control qubit $\mathbf{C}$ that determines their causal order. The combined action is represented by Kraus operators
\begin{equation}
    K_{ij}:= B_j A_i \otimes \ket{0}_{\rm c}\bra{0} + A_i B_j \otimes \ket{1}_{\rm c}\bra{1},
\end{equation}
where $\{A_i\}_i$ and $\{B_j\}_j$ are the Kraus operators of $\mathcal{A}$ and $\mathcal{B}$, respectively, and $\{\ket{0}_{\rm c}, \ket{1}_{\rm c}\}$ form the computational basis of the controller. Therefore, denoting the initial state of the target and control systems as $\varrho_{\rm in}$, the resulting map produced by the quantum SWITCH outputs a state
\begin{equation}\label{app:SwitchGeneral}
   \mathcal{S}(\mathcal{A},\mathcal{B})[\varrho_{\rm in}] = \sum_{ij}K_{ij}\varrho_{\rm in}K_{ij}^{\dagger}.
\end{equation}

We now specialize to the configuration introduced in Section~\ref{sec:HeatEngineSWITCH}, where the target system is the working medium $\mathbf{Q}$ and the controller is $\mathbf{C}$. The initial joint state $\varrho_{\rm in}$ satisfies the local thermal condition~\eqref{eq:localThermalState}, and the two operations are the measurement channels associated with apparata $\mathbf{A}$ and $\mathbf{A}'$, i.e., $\mathcal{A} = \mathcal{M}^a$ and $\mathcal{B} = \mathcal{M}^{a'}$ (see Fig.~\ref{fig:switchSetup}). Because $\mathcal{M}^a$ maps any input state to $\rho_a := \operatorname{diag}[a, 1-a]$, a straightforward calculation shows that the SWITCH of these two channels,
$\mathcal{S}^{a,a'} := \mathcal{S}(\mathcal{M}^{a}, \mathcal{M}^{a'})$, transforms $\varrho_{\rm in}$ according to
\begin{widetext}
\begin{equation}\label{app:SwitchGeneralMeasChan}
   \mathcal{S}^{a,a'}[\varrho_{\rm in}] = \lambda[\varrho_{\rm in}] \rho_{a'} \otimes \ket{0}_{\rm c}\bra{0} + (1 - \lambda[\varrho_{\rm in}]) \rho_a \otimes \ket{1}_{\rm c}\bra{1} + \Lambda_X^{(a,a')}[\varrho_{\rm in}] \otimes X_{\rm c} + \Lambda_Y^{(a,a')}[\varrho_{\rm in}] \otimes Y_{\rm c},
\end{equation}
\end{widetext}
where
\begin{eqnarray}\label{eq:app:lambdaDef}
    \lambda[\varrho_{\rm in}] &:=& \frac{1}{2}\bigl(1 + \operatorname{tr}[\varrho_{\rm in} Z_{\rm c}]\bigr), \\
    \Lambda_X^{(a,a')}[\varrho_{\rm in}] &:=& \frac{1}{2}\Bigl(\rho_{a'} \operatorname{tr}_{\rm c}[\varrho_{\rm in} \ket{1}_{\rm c}\bra{0}] \rho_{a} + \mathrm{h.c.}\Bigr), \\
    \Lambda_Y^{(a,a')}[\varrho_{\rm in}] &:=& \frac{i}{2}\Bigl(\rho_{a'} \operatorname{tr}_{\rm c}[\varrho_{\rm in} \ket{1}_{\rm c}\bra{0}] \rho_{a} - \mathrm{h.c.}\Bigr).
\end{eqnarray}
The terms $\Lambda_X^{(a,a')}$ and $\Lambda_Y^{(a,a')}$ capture the non-classical contributions arising from placing the measurements in SCO.

\textit{Incoherent mode.}
In the incoherent mode of the quantum SWITCH, the control qubit $\mathbf{C}$ is left unobserved. Tracing it out yields a probabilistic mixture of the two possible causal orders, with probabilities $\{\lambda[\varrho_{\rm in}], 1 - \lambda[\varrho_{\rm in}]\}$. If the control and the working medium share initial correlations, the state after the quantum SWITCH corresponds to the mixture of the system states once the state of the control is known,
\begin{align}
        \rho_{\rm inc}^{(1)}&= \operatorname{tr}_{c}[\mathcal{S}^{a,a'}[\varrho_{\rm in}]] \nonumber \\
        &= \lambda[\varrho_{\rm in}](\mathcal{M}^{a'}\circ\mathcal{M}^a)[\rho^{(0)}_{0}] \nonumber \\
        &\qquad +(1-\lambda[\varrho_{\rm in}])(\mathcal{M}^a\circ\mathcal{M}^{a'})[\rho^{(0)}_{1}],
\end{align}
where $\lambda[\varrho_{\rm in}]$ is defined in \eqref{eq:app:lambdaDef}, and
\begin{equation}
    \rho^{(0)}_{\ell}:=\frac{\operatorname{tr}_{\rm c}[\varrho_{\rm in} \ket{\ell}_{\rm c}\bra{\ell}]}{\operatorname{tr}[\varrho_{\rm in} \ket{\ell}_{\rm c}\bra{\ell}]},
\end{equation}
where $\ell \in \{0,1\}$. Given the structure of the chosen channel,
\begin{equation}
    \mathcal{M}^{a'}\circ\mathcal{M}^a = \mathcal{M}^{a'}, \qquad \mathcal{M}^a\circ\mathcal{M}^{a'}= \mathcal{M}^a,
\end{equation}
this follows from \eqref{eq:MeasChan}, since each map $\mathcal{M}^{\lambda}$ outputs the fixed diagonal state $\operatorname{diag}[\lambda, 1-\lambda]$, so any subsequent channel $\mathcal{M}^{\lambda'}$ overwrites the action of the previous one. Therefore, the resulting map effectively implements a single measurement channel $\mathcal{M}^{\bar{a}^{\mathrm{inc}}}$ with the averaged strength
\begin{equation}\label{app:incStrength}
\bar{a}^{\mathrm{inc}} = \lambda[\varrho_{\rm in}] a' + \bigl(1 - \lambda[\varrho_{\rm in}]\bigr) a.
\end{equation}
The post-measurement state of the working medium is then
\begin{equation}\label{app:incState}
    \rho^{(1)}_{\rm inc} = \rho_{\bar{a}^{\rm inc}}.
\end{equation}

\textit{Coherent mode.}
In the coherent mode, the control qubit $\mathbf{C}$ is measured in the basis $\{|\pm\rangle_{\rm c}\}$, defined as $|\pm\rangle_{\rm c} := \frac{1}{\sqrt{2}}(|0\rangle_{\rm c} \pm e^{i \phi'} |1\rangle_{\rm c})$ for some phase $\phi' \in [0, 2\pi]$. Each measurement outcome ($|+\rangle_{\rm c}$ or $|-\rangle_{\rm c}$) occurs with probability
\begin{equation}
    p^{(\pm)} = \operatorname{tr}[\mathcal{S}(\mathcal{M}^{a}, \mathcal{M}^{a'})[\varrho_{\rm in}](\mathbb{I} \otimes |\pm\rangle_{\rm c}\langle\pm|)].
\end{equation}
Using \eqref{app:SwitchGeneralMeasChan}, we obtain:
\begin{align}\label{app:probability}
    p^{(\pm)} &= \frac{1}{2}\Bigl( \operatorname{tr}[\mathcal{S}(\mathcal{M}^{a}, \mathcal{M}^{a'})[\varrho_{\rm in}]] \nonumber \\
    &\qquad \pm \operatorname{tr}[\mathcal{S}(\mathcal{M}^{a}, \mathcal{M}^{a'})[\varrho_{\rm in}](\mathbb{I} \otimes e^{-i\phi'}|0\rangle_{\rm c}\langle 1| \nonumber \\
    &\qquad + \mathbb{I} \otimes e^{i\phi'}|1\rangle_{\rm c}\langle 0|)]\Bigr) \nonumber \\
    &= \frac{1}{2} \Bigl(1 \pm \operatorname{tr}\Bigl[\Xi_{\phi'}^{(a,a')}[\varrho_{\rm in}]\Bigr] \Bigr),
\end{align}
where
\begin{equation}
     \Xi_{\phi'}^{(a,a')}[\varrho_{\rm in}] =  2\bigl(\cos(\phi') \Lambda_X^{(a,a')}[\varrho_{\rm in}]+ \sin(\phi') \Lambda_Y^{(a,a')}[\varrho_{\rm in}]\bigr).
\end{equation}
Therefore, each measurement outcome prepares the corresponding conditional state
\begin{eqnarray}
     \rho^{(1)}_{\pm} &=& \operatorname{tr}_{c}[\mathcal{S}^{a,a'}[\varrho_{\rm in}]|\pm\rangle_{\rm c}\langle\pm|] \nonumber \\
     \label{app:postselectedstates} &=& \frac{1}{2p^{(\pm)}} \Bigl(\rho_{\bar{a}^{\rm inc}} \pm \Xi_{\phi'}^{(a,a')}[\varrho_{\rm in}] \Bigr),
\end{eqnarray}
with probability $p^{(\pm)}$.

The operators $\Lambda_{X/Y}^{(a,a')}[\varrho_{\rm in}]$ can introduce quantum coherence in the states $\rho^{(1)}_{\pm}$. The corresponding off-diagonal elements,
\begin{eqnarray}
\rho^{(1)}_{01, \pm} = \pm \frac{1}{2p^{(\pm)}}\operatorname{tr}\Bigl[
\Xi_{\phi'}^{(a,a')}[\varrho_{\rm in}]\ket{1}\bra{0} \Bigr],
\end{eqnarray}
are determined by
\begin{widetext}
\begin{eqnarray}
    \operatorname{tr}[\Xi_{\phi'}^{(a,a')}[\varrho_{\rm in}]\ket{1}\bra{0}] &=& 2\operatorname{tr}[(\cos(\phi') \Lambda_X^{(a,a')}[\varrho_{\rm in}] + \sin(\phi') \Lambda_Y^{(a,a')}[\varrho_{\rm in}]) \ket{1}\bra{0}] \nonumber \\
    &=& e^{i\phi'} a'(1-a) \operatorname{tr}\Bigl[\varrho_{\rm in} (\ket{1}\bra{0} \otimes \ket{1}_{\rm c}\bra{0} ) \Bigr] + e^{-i\phi'} a(1-a') \operatorname{tr}\Bigl[\varrho_{\rm in} (\ket{1}\bra{0} \otimes \ket{0}_{\rm c}\bra{1} ) \Bigr].
\end{eqnarray}
\end{widetext}
Unlike in the case of definite or probabilistic causal order, these coherence terms do not necessarily vanish, depending explicitly on the off-diagonal blocks of $\varrho_{\rm in}$. If, however, $\Lambda_{X/Y}^{(a,a')}[\varrho_{\rm in}]$ are diagonal, preparation of the states \eqref{app:postselectedstates} merely corresponds to applying one of two effective measurement channels with strengths
\begin{equation}\label{app:eq:postSelStrength}
    \bar{a}_{\pm} = \frac{1}{2p^{(\pm)}} (\bar{a}^{\rm inc} \pm \Delta^{(a,a')}),
\end{equation}
where
\begin{align}\label{app:eq:postSelStrengthDelta}
    \Delta^{(a,a')} &= \operatorname{tr}[\Xi_{\phi'}^{(a,a')}[\varrho_{\rm in}]\ket{0}\bra{0}] \nonumber \\
    &= aa'\operatorname{tr}\Bigl[\varrho_{\rm in} \Bigl(\ket{0}\bra{0} \otimes (\cos(\phi') X_{\rm c} + i\sin(\phi') Y_{\rm c})\Bigr) \Bigr],
\end{align}
Nevertheless, unlike \eqref{app:incStrength}, the effective measurement strengths $\bar{a}_{\pm}$ need not lie within the interval $[\min(a,a'),\max(a,a')]$, underscoring that the SCO leads to operational outcomes unattainable by any classical mixture of the two measurement sequences.

\subsection{Energy and entropy in the extended heat engine}\label{app:varEnerEntrSwitch}

Before the heat engine starts, the working medium $\mathbf{Q}$ is locally thermal, described by the Gibbs state~\eqref{app:eq:GibbsState} at inverse temperature $\beta$. Consequently, its initial internal energy and entropy coincide with those in the definite-order model discussed in Appendix~\ref{app:heatEngVar}, given by \eqref{app:eq:zeroEnergy} and~\eqref{app:eq:zeroEntropy}, respectively.

\emph{Incoherent mode.}
In the incoherent mode, the quantum SWITCH effectively applies a single measurement channel $\mathcal{M}^{\bar{a}^{\mathrm{inc}}}$ to $\mathbf{Q}$ in the first stroke. The extended engine is therefore equivalent to the definite-order model with measurement strength $a = \bar{a}^{\mathrm{inc}}$. The corresponding energy exchanges are
\begin{eqnarray}
    \mathcal{Q}_{\rm hot}^{\rm inc} &=& \varepsilon \Bigl( 1 - 2\bar{a}^{\rm inc} +  \tanh(\beta\varepsilon) \Bigr), \\
    \mathcal{W}_{\rm ext}^{\rm inc} &=& \varepsilon \Bigl( 1-2\bar{a}^{\rm inc} + |1-2\bar{a}^{\rm inc}| \Bigr), \\
    \mathcal{Q}_{\rm cold}^{\rm inc} &=& \varepsilon \Bigl( |1-2\bar{a}^{\rm inc}| - \tanh(\beta\varepsilon)\Bigr),
\end{eqnarray}
subject to the constraints
\begin{eqnarray}
    \bar{a}^{\rm inc} &>& \frac{1}{2}\Bigl( 1 - \tanh(\beta\varepsilon) \Bigr), \\
    \bar{a}^{\rm inc} &<& \frac{1}{2}\Bigl( 1 + \tanh(\beta\varepsilon) \Bigr), \\
    b^{\rm inc} &=& \frac{1}{2}\Bigl( 1 + |1-2\bar{a}^{\rm inc}| \Bigr).
\end{eqnarray}

\begin{table*}[ht!]
\centering
\caption{Variations of internal energy and entropy during each stroke of the extended heat engine operating in the coherent mode of the quantum SWITCH, conditioned on the measurement outcome ($|\pm\rangle_{\rm c}$) of the controller $\mathbf{C}$.}%
\label{tab:energy2}
\begin{tabular}{|l|ll|ll|}
    \hline
    &\multicolumn{2}{l}{Internal energy variation}&\multicolumn{2}{|l|}{Entropy variation} \\
    \hline
    \textbf{Stroke 1:} & $\Delta U^{(1)}_{\pm}$ &$= \varepsilon (1 - 2\bar{a}_{\pm} + \tanh(\beta \varepsilon))$ & $\Delta S^{(1)}_{\pm}$ &$= \mathcal{H}_\text{bin}\Bigl(\frac{1+\sqrt{(1-2\bar{a}_{\pm})^2 + 4|\rho_{01, \pm}^{(1)}|^2}}{2}\Bigr) - \mathcal{H}_\text{bin}\left(\frac{1 + \tanh(\beta \varepsilon)}{2}\right)$ \\
    \textbf{Stroke 2:} & $\Delta U^{(2)}_{\pm}$ &$= -\varepsilon\Bigl(1-2\bar{a}_{\pm} + \sqrt{(1-2\bar{a}_{\pm})^2 + 4|\rho_{01, \pm}^{(1)}|^2}\Bigr)$ & $\Delta S^{(2)}_{\pm}$ &$= 0$ \\
    \textbf{Stroke 3:} & $\Delta U^{(3)}_{\pm}$ &$= \varepsilon \Bigl( \sqrt{(1-2\bar{a}_{\pm})^2 + 4|\rho_{01, \pm}^{(1)}|^2} - \tanh(\beta\varepsilon)\Bigr)$ & $\Delta S^{(3)}_{\pm}$ &$= \mathcal{H}_\text{bin}\left(\frac{1 + \tanh(\beta \varepsilon)}{2}\right) - \mathcal{H}_\text{bin}\Bigl(\frac{1+\sqrt{(1-2\bar{a}_{\pm})^2 + 4|\rho_{01, \pm}^{(1)}|^2}}{2}\Bigr)$ \\
    \hline
\end{tabular}
\end{table*}

\emph{Coherent mode: First stroke.}
In the coherent mode, the first stroke applies the two measurements $\mathcal{M}^a$ and $\mathcal{M}^{a'}$ in a superposition of causal orders, leaving $\mathbf{Q}$ in one of the conditional states~\eqref{app:postselectedstates}. These states may carry coherence, contributing to their entropy. Their internal energy and von Neumann entropy are
\begin{eqnarray}
    U^{(1)}_{\pm} &=& \varepsilon (1 - 2\bar{a}_{\pm}), \\
    S^{(1)}_{\pm} &=& \mathcal{H}_{\mathrm{bin}}\Bigl(\frac{1}{2} \bigl( 1 + \sqrt{(1-2\bar{a}_{\pm})^2 + 4|\rho_{01, \pm}^{(1)}|^2}\bigr)\Bigr),
\end{eqnarray}
Accordingly, the first stroke produces variations
\begin{eqnarray}
    \Delta U^{(1)}_{\pm} &=& \varepsilon (1 - 2\bar{a}_{\pm} + \tanh(\beta\varepsilon)), \\
    \Delta S^{(1)}_{\pm} &=& \mathcal{H}_{\mathrm{bin}}\Bigl(\frac{1}{2} \bigl( 1 + \sqrt{(1-2\bar{a}_{\pm})^2 + 4|\rho_{01, \pm}^{(1)}|^2}\bigr)\Bigr) \nonumber \\
    &-& \mathcal{H}_{\mathrm{bin}}\Bigl(\frac{1}{2} \bigl( 1 + \tanh(\beta\varepsilon)\bigr)\Bigr).
\end{eqnarray}
For this step to correspond to heat absorption from the measurement devices $\mathbf{A}$ and $\mathbf{A}'$, we impose $\mathcal{Q}_{\rm hot}^{(\pm)} = \Delta U^{(1)}_{\pm} \geq 0$ and $\Delta S^{(1)}_{\pm} > 0$. These lead to the constraints
\begin{eqnarray}
    1 - 2\bar{a}_{\pm} + \tanh(\beta\varepsilon) &\geq& 0, \\
    \sqrt{(1-2\bar{a}_{\pm})^2 + 4|\rho_{01, \pm}^{(1)}|^2} &<& \tanh(\beta\varepsilon),
\end{eqnarray}
which, in turn, yield
\begin{eqnarray}
    \label{app:eq:aLowBoundSwitch} \bar{a}_{\pm} &>& \frac{1}{2}\bigl(1 - \tanh(\beta\varepsilon)\bigr), \\
    \label{app:eq:aUppBoundSwitch} \bar{a}_{\pm} &<& \frac{1}{2}\bigl(1 + \tanh(\beta\varepsilon)\bigr), \\
    \label{app:eq:cohBoundSwitch} |\rho_{01, \pm}^{(1)}| &<& \frac{1}{2}\sqrt{\tanh^2(\beta\varepsilon) - (1-2\bar{a}_{\pm})^2}.
\end{eqnarray}
Taking into account \eqref{app:eq:postSelStrength}, we have
\begin{equation}
    \sum_{\pm} p^{(\pm)}\bar{a}_{\pm} = \bar{a}^{\rm inc},
\end{equation}
which implies that the average heat absorbed during the first stroke equals that in the incoherent mode,
\begin{eqnarray}
    \langle \mathcal{Q}_{\rm hot}^{\rm coh} \rangle &=& \sum_{\pm} p^{(\pm)} \mathcal{Q}_{\rm hot}^{(\pm)} \nonumber \\
    &=& \varepsilon (1 - 2\bar{a}^{\rm inc} + \tanh(\beta\varepsilon)) \nonumber \\
    &\equiv& \mathcal{Q}_{\rm hot}^{\rm inc}.
\end{eqnarray}

\emph{Coherent mode: Second stroke.}
In the isentropic second stroke, work is extracted from the working medium $\mathbf{Q}$ by the measurement apparatus $\mathbf{B}$. The corresponding strength $b=b_\pm$ must enforce $\Delta S^{(2)}_\pm=0$, which implies
\begin{equation}
    |1-2b_{\pm}| = \sqrt{(1-2\bar{a}_{\pm})^2 + 4|\rho_{01, \pm}^{(1)}|^2}.
\end{equation}
Accordingly,
\begin{eqnarray}\label{U3theta}
    U^{(2)}_{\pm} &=& \Gamma^{(2)}_{\pm} \sqrt{(U^{(1)}_{\pm})^2 + 4\varepsilon^2|\rho_{01, \pm}^{(1)}|^2},
\end{eqnarray}
where $\Gamma^{(2)}_{\pm} \in \{-1,1\}$ selects the isentropic branch for the outcome $|\pm\rangle_{\rm c}$ of $\mathbf{C}$. To extract work, we require $U^{(2)}_{\pm} \leq U^{(1)}_{\pm}$, so that $\mathcal{W}^{(\pm)}_{\rm ext} = -\Delta U^{(2)}_{\pm}$, yielding
\begin{align}
    \mathcal{W}_{\rm ext}^{(\pm)} &= \varepsilon \Bigl( U^{(1)}_{\pm} - \Gamma^{(2)}_{\pm} \sqrt{(U^{(1)}_{\pm})^2 + 4\varepsilon^2|\rho_{01,\pm}^{(1)}|^2} \Bigr) \nonumber \\
    \label{app:eq:extWork} &= \varepsilon \Bigl( 1-2\bar{a}_{\pm} - \Gamma^{(2)}_{\pm} \sqrt{(1-2\bar{a}_{\pm})^2 + 4|\rho_{01, \pm}^{(1)}|^2} \Bigr), \\
    b_{\pm} &= \frac{1}{2}\Bigl( 1 - \Gamma^{(2)}_{\pm}\sqrt{(1-2\bar{a}_{\pm})^2 + 4|\rho_{01, \pm}^{(1)}|^2} \Bigr).
\end{align}
Imposing $\mathcal{W}^{(\pm)}_{\rm ext} \geq 0$ fixes $\Gamma^{(2)}_{\pm} = -1$, and we obtain
\begin{eqnarray}
    \mathcal{W}_{\rm ext}^{(\pm)} &=& \varepsilon \Bigl( 1-2\bar{a}_{\pm} + \sqrt{(1-2\bar{a}_{\pm})^2 + 4|\rho_{01,\pm}^{(1)}|^2} \Bigr), \nonumber \\
    b_{\pm} &=& \frac{1}{2}\Bigl( 1 + \sqrt{(1-2\bar{a}_{\pm})^2 + 4|\rho_{01, \pm}^{(1)}|^2} \Bigr).
\end{eqnarray}
This shows that, unlike in the definite-order engine of Appendix~\ref{app:heatEngVar} or the incoherent mode, positive work can be extracted even when the population imbalance alone would not allow it (e.g., for $\bar{a}_{\pm} > 1/2$), provided the first stroke generates coherence, $|\rho_{01,\pm}^{(1)}| > 0$. The average extracted work in the coherent mode of the quantum SWITCH is
\begin{widetext}
\begin{equation}
     \langle \mathcal{W}_{\rm ext}^{\rm coh} \rangle = \sum_{\pm} p^{(\pm)} \mathcal{W}_{\rm ext}^{(\pm)} = \varepsilon \Bigl( 1-2\bar{a}^{\rm inc} + \sum_{\pm} p^{(\pm)} \sqrt{(1-2\bar{a}_{\pm})^2 + 4|\rho_{01,\pm}^{(1)}|^2} \Bigr).
\end{equation}
\end{widetext}

\emph{Coherent mode: Third stroke.}
Subsequent thermalization with the cold reservoir $\mathbf{R}$ returns $\mathbf{Q}$ to $\rho^{(3)}_{\pm} = \rho^{(0)}$, so $U^{(3)}_{\pm} = U^{(0)}$ and $S^{(3)}_{\pm} = S^{(0)}$. Hence,
\begin{eqnarray}
    \Delta U^{(3)}_{\pm} &=& -\varepsilon \Bigl( \tanh(\beta\varepsilon) + 1 - 2b_{\pm}\Bigr), \\
    \Delta S^{(3)}_{\pm} &=& \mathcal{H}_{\mathrm{bin}} \Bigl(\frac{1}{2}(1+\tanh(\beta\varepsilon))\Bigr) - \mathcal{H}_{\mathrm{bin}} (b_{\pm}).
\end{eqnarray}
Similarly to the first stroke, the energy exchange is interpreted as heat $\mathcal{Q}_{\rm cold}^{(\pm)}$, yet transferred from the working medium $\mathbf{Q}$ to the reservoir $\mathbf{R}$. To ensure the correct heat flow direction, we impose the condition $\mathcal{Q}_{\rm cold}^{(\pm)} = \Delta U^{(3)}_{\pm} \leq 0$, which yields
\begin{equation}
    \mathcal{Q}_{\rm cold}^{(\pm)} = \varepsilon \Bigl( \sqrt{(1-2\bar{a}_{\pm})^2 + 4|\rho_{01, \pm}^{(1)}|^2} - \tanh(\beta\varepsilon)\Bigr),
\end{equation}
subject to the constraints \eqref{app:eq:aLowBoundSwitch}--\eqref{app:eq:cohBoundSwitch} on $\bar{a}_{\pm}$ and $\rho_{01, \pm}^{(1)}$. The three strokes are summarized in Table~\ref{tab:energy2}.

\begin{figure}[t]
    \centering
    \includegraphics[width=0.9\columnwidth]{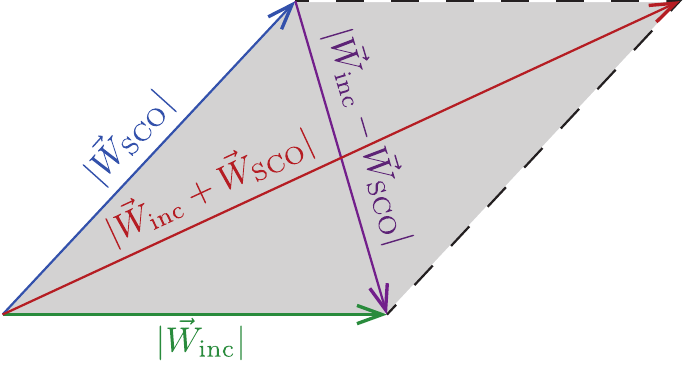}
    \caption[Geometric interpretation of the SCO contribution to the extracted work in the extended heat engine.]{\textbf{Geometric interpretation of the SCO contribution to the extracted work in the extended heat engine.} The contributions of the incoherent and coherent regimes of the quantum SWITCH to the extracted work are represented by the two-dimensional vectors defined in \eqref{eq:incVector} and \eqref{eq:cohVector}. The components of these vectors correspond, respectively, to the work arising from population imbalance (associated with permutations of the diagonal elements of the working medium state) and from the consumption of quantum coherence. Geometrically, the vectors \eqref{eq:incVector} and \eqref{eq:cohVector} span a parallelogram in the work space, where $\vec{W}_{\rm inc}$, having only a nonzero first component, serves as the base. The SCO contribution $\Delta \eta^{\rm coh}$ to the efficiency can then be interpreted as the excess magnitude, given by the difference between the average length of the two diagonals of the parallelogram and the length of its base, reflecting the enhancement in extractable work due to SCO.}
    \label{fig:geomWork}
\end{figure}

A natural question concerns constraints on the engine's efficiency and the explicit contribution of SCO. From \eqref{app:probability} and \eqref{app:eq:postSelStrength} one finds
\begin{eqnarray}
     p^{(\pm)}(1-2\bar{a}_\pm) &=& p^{(\pm)} - \bar{a}^{\rm inc} \mp\Delta^{(a,a')} \nonumber \\
    &=& \frac{1}{2} \Bigl(1 - 2\bar{a}^{\rm inc} \mp \operatorname{tr}\Bigl[\Xi_{\phi'}^{(a,a')}[\varrho_{\rm in}]Z\Bigr]\Bigr). \nonumber
\end{eqnarray}
This identity allows us to express the efficiency $\tilde{\eta}^{\rm coh} = \frac{1}{2}(\tilde{\eta}^{\rm coh}_+ + \tilde{\eta}^{\rm coh}_-)$ and the SCO contribution $\Delta\eta^{\rm coh} = \frac{1}{2}(\Delta\eta^{\rm coh}_+ + \Delta\eta^{\rm coh}_-)$, with
\begin{widetext}
\begin{eqnarray}
    \tilde{\eta}_\pm^{\rm coh} &=& \frac{\sqrt{\Bigl(1-2\bar{a}^{\rm inc} \pm \operatorname{tr}\Bigl[\Xi_{\phi'}^{(a,a')}[\varrho_{\rm in}] Z\Bigr]\Bigr)^2 + 4\Bigl|\operatorname{tr}\Bigl[\Xi_{\phi'}^{(a,a')}[\varrho_{\rm in}]|1\rangle\langle 0|\Bigr]\Bigr|^2} + 1-2\bar{a}^{\rm inc}}{1-2\bar{a}^{\rm inc}+\tanh(\beta\varepsilon)}, \\
    \Delta \eta^{\rm coh}_\pm &=& \frac{\sqrt{\Bigl(1-2\bar{a}^{\rm inc} \pm \operatorname{tr}\Bigl[\Xi_{\phi'}^{(a,a')}[\varrho_{\rm in}] Z\Bigr]\Bigr)^2 + 4\Bigl|\operatorname{tr}\Bigl[\Xi_{\phi'}^{(a,a')}[\varrho_{\rm in}]|1\rangle\langle 0|\Bigr]\Bigr|^2} - |1-2\bar{a}^{\rm inc}|}{1-2\bar{a}^{\rm inc}+\tanh(\beta\varepsilon)}. \label{app:eq:delta_eta_coh}
\end{eqnarray}
\end{widetext}
Likewise, the efficiencies $\eta^{(\pm)}$ associated with measurement outcomes of $\mathbf{C}$ read
\begin{widetext}
\begin{equation}\label{eq:effPostSelectedGen}
    \eta^{(\pm)} = \frac{\sqrt{\Bigl(1-2\bar{a}^{\rm inc} \mp \operatorname{tr}\Bigl[\Xi_{\phi'}^{(a,a')}[\varrho_{\rm in}] Z\Bigr]\Bigr)^2 + 4\Bigl|\operatorname{tr}\Bigl[\Xi_{\phi'}^{(a,a')}[\varrho_{\rm in}]|1\rangle\langle 0|\Bigr]\Bigr|^2} + 1-2\bar{a}^{\rm inc} \mp \operatorname{tr}\Bigl[\Xi_{\phi'}^{(a,a')}[\varrho_{\rm in}] Z\Bigr]}{1-2\bar{a}^{\rm inc} \mp \operatorname{tr}\Bigl[\Xi_{\phi'}^{(a,a')}[\varrho_{\rm in}] Z\Bigr] + \Bigl(1 \pm \operatorname{tr}\Bigl[\Xi_{\phi'}^{(a,a')}[\varrho_{\rm in}] \Bigr]\Bigr)\tanh(\beta\varepsilon)}.
\end{equation}
\end{widetext}
A convenient geometric form of \eqref{app:eq:delta_eta_coh} is
\begin{equation}\label{app:eq:delta_eta_cohGen}
    \Delta \eta^{\rm coh} = \frac{\frac{1}{2}\Bigl(|\vec{W}_{\rm inc} + \vec{W}_{\rm SCO}|+|\vec{W}_{\rm inc} - \vec{W}_{\rm SCO}|\Bigr) - |\vec{W}_{\rm inc}|}{1-2\bar{a}^{\rm inc}+\tanh(\beta\varepsilon)},
\end{equation}
where the two-dimensional vectors
\begin{eqnarray}
    \vec{W}_{\rm inc} &=& \Bigl(1-2\bar{a}^{\rm inc} , \; 0 \Bigr) \nonumber \\
    &:=& (W^{\rm diag}_{\rm inc}, \; W^{\rm off}_{\rm inc}) \label{eq:incVector},
\end{eqnarray}
and
\begin{eqnarray}
    \vec{W}_{\rm SCO} &=& \Bigl(\operatorname{tr}\Bigl[\Xi_{\phi'}^{(a,a')}[\varrho_{\rm in}] Z\Bigr] , \; 2\Bigl|\operatorname{tr}\Bigl[\Xi_{\phi'}^{(a,a')}[\varrho_{\rm in}]|1\rangle\langle 0|\Bigr]\Bigr| \Bigr) \nonumber \\
    &:=& (W^{\rm diag}_{\rm SCO}, \; W^{\rm off}_{\rm SCO}) \label{eq:cohVector},
\end{eqnarray}
capture the contributions of the incoherent and coherent modes, respectively, and separate the diagonal (population-imbalance) and off-diagonal (coherence) contributions to extractable work. By the triangle inequality, $\Delta\eta^{\rm coh} \geq 0$: interference between causal orders cannot reduce the efficiency (see Fig.~\ref{fig:geomWork}).

Physically, $W^{\rm diag}$ quantifies work available from redistributing energy populations, whereas $W^{\rm off}$ captures work available from consuming coherence. In the incoherent mode of the quantum SWITCH, the state is diagonal, so that $W^{\rm off}_{\rm inc}=0$, and work arises solely from population permutations. In the coherent mode, the conditional states \eqref{app:postselectedstates} may carry coherence: if
\begin{equation}\label{eq:cohPsState}
    \operatorname{tr}\Bigl[\Xi_{\phi'}^{(a,a')}[\varrho_{\rm in}]\ket{1}\bra{0}\Bigr] \neq 0,
\end{equation}
then $W^{\rm off}_{\rm SCO}>0$, and the coherent mode enhances the efficiency of the heat engine by $\Delta\eta^{\rm coh} > 0$. The same condition guarantees non-zero efficiencies $\eta^{(\pm)}$ associated with outcomes of a measurement of $\mathbf{C}$ as well.

If the conditional states \eqref{app:postselectedstates} remain diagonal ($W^{\rm off}_{\rm SCO}=0$), the coherent mode effectively reduces to measurement channels of strengths $\bar{a}_\pm$, and the SCO contribution becomes
\begin{widetext}
\begin{eqnarray}
    \Delta \eta^{\rm coh}_{W^{\rm off}_{\rm SCO} = 0} &=& \frac{\frac{1}{2}\Bigl|1-2\bar{a}^{\rm inc} + \operatorname{tr}\Bigl[\Xi_{\phi'}^{(a,a')}[\varrho_{\rm in}] Z\Bigr]\Bigr| + \frac{1}{2}\Bigl|1-2\bar{a}^{\rm inc} - \operatorname{tr}\Bigl[\Xi_{\phi'}^{(a,a')}[\varrho_{\rm in}] Z\Bigr]\Bigr| - |1-2\bar{a}^{\rm inc}|}{1-2\bar{a}^{\rm inc}+\tanh(\beta\varepsilon)}.
\end{eqnarray}
\end{widetext}
It is strictly positive whenever
\begin{equation}\label{eq:nonZeroEffCond}
    |1-2\bar{a}^{\rm inc}| < \Bigl| \operatorname{tr}[\Xi_{\phi'}^{(a,a')}[\varrho_{\rm in}] Z] \Bigr|,
\end{equation}
so that
\begin{equation}\label{eq:nonZeroEffCondGen}
    \Bigl|W_{\rm inc}^{\rm diag}\Bigr| < \Bigl| W_{\rm SCO}^{\rm diag} \Bigr|,
\end{equation}
and given by
\begin{eqnarray}
    \Delta \eta^{\rm coh}_{W^{\rm off}_{\rm SCO} = 0} &=& \frac{\Bigl|\operatorname{tr}\Bigl[\Xi_{\phi'}^{(a,a')}[\varrho_{\rm in}] Z\Bigr]\Bigr| - |1-2\bar{a}^{\rm inc}|}{1-2\bar{a}^{\rm inc}+\tanh(\beta\varepsilon)}.
\end{eqnarray}
That is, whenever the population imbalance of the state produced in the incoherent mode of the quantum SWITCH is smaller than the population-imbalance variation induced by SCO in the coherent mode.

\begin{figure}[t]
    \centering
    \includegraphics[width=\columnwidth]{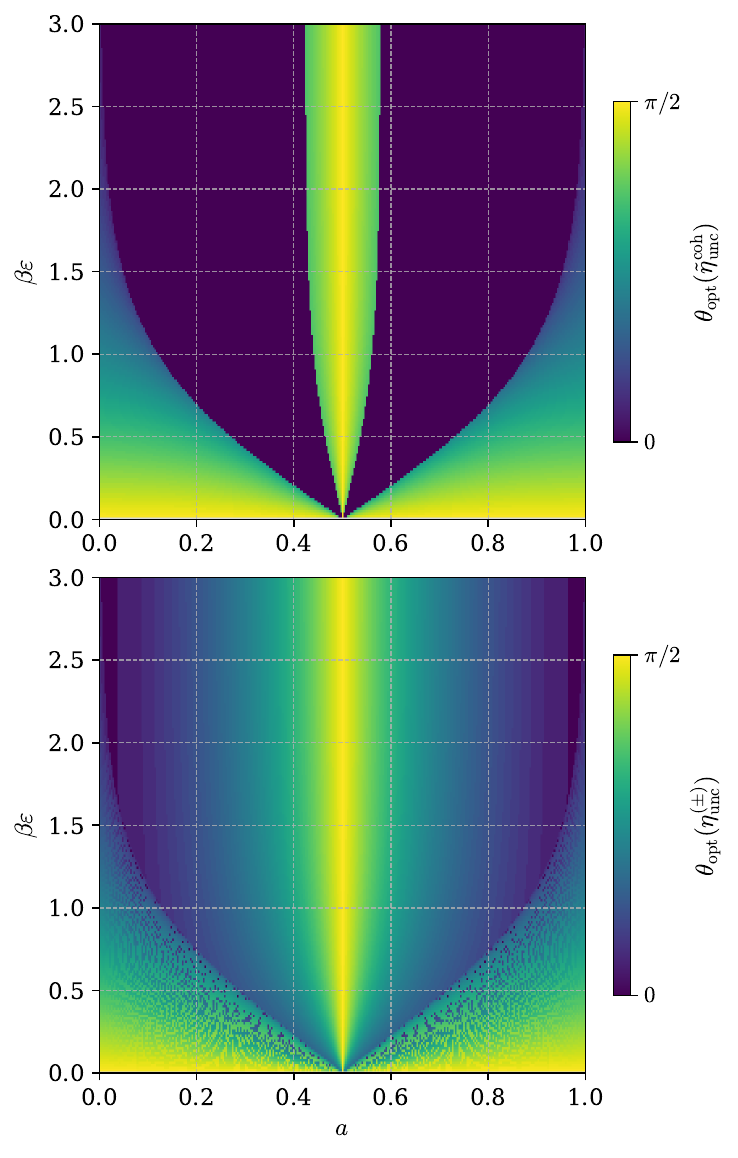}
    \caption[Optimal angles $\theta$ maximizing the efficiencies $\tilde{\eta}^{\rm coh}$ and $\eta^{(\pm)}$ for an uncorrelated initial state.]{\textbf{Optimal angles $\theta$ maximizing the efficiencies $\tilde{\eta}^{\rm coh}$ and $\eta^{(\pm)}$ for an uncorrelated initial state.} The plots show the optimal measurement angle $\theta_{\rm opt}$ that maximizes the efficiency of the heat engine as a function of the measurement strength and the inverse temperature, for an uncorrelated initial state $\varrho_{\rm in}^{\rm unc}$ and measurement apparata of complementary strengths $a$ and $a' = 1-a$. The upper panel presents $\theta_{\rm opt}$ corresponding to the efficiency $\tilde{\eta}^{\rm coh}$, while the lower panel presents $\theta_{\rm opt}$ for the efficiencies $\eta^{(\pm)}$.}
    \label{fig:optangle_unc}
\end{figure}


\section{Impact of initial correlations}

\subsection{Uncorrelated initial state} \label{app:uncorrelated}

In the usual setting of the quantum SWITCH \cite{rozema2024experimental}, $\mathbf{Q}$ and $\mathbf{C}$ initially share no correlations, so that $\varrho_{\rm in} = \rho^{(0)} \otimes \omega$ \cite{dieguez2023thermal}. In turn, the state of $\mathbf{C}$ can be parameterized as
\begin{equation}
\omega = \zeta \ketbrac{\psi_{\theta,\phi}}+ (1-\zeta) \ketbrac{\psi_{\theta-\pi,\phi}},
\end{equation}
where
\begin{equation}\label{eq:pureStateUnc}
    \ket{\psi_{\theta,\phi}}_{\rm c} = \cos\left(\frac{\theta}{2}\right)\ket{0}_{\rm c}+\sin\left(\frac{\theta}{2}\right)e^{i\phi}\ket{1}_{\rm c},
\end{equation}
and $\theta \in [0, \pi]$ and $\phi \in [0, 2\pi]$ are the Bloch-sphere angles, whereas $\zeta \in [0,1]$ encodes the purity of $\omega$: in particular, $\omega$ is pure if $\zeta=1$. Applying \eqref{app:SwitchGeneralMeasChan}, we obtain the output of the map produced by the quantum SWITCH,
\begin{eqnarray}\label{app:SwitchGeneralMeasChanUnc}
    \lambda[\rho^{(0)} \otimes \omega] &=& \frac{1}{2}\Bigl(1 + (2\zeta-1)\cos(\theta)\Bigr), \\
    \Lambda_X^{(a,a')}[\rho^{(0)} \otimes \omega] &=& \frac{1}{2}\left(2\zeta-1 \right) \cos(\phi) \sin(\theta) \rho_{a'} \rho^{(0)} \rho_{a}, \\
    \Lambda_Y^{(a,a')}[\rho^{(0)} \otimes \omega] &=& \frac{1}{2}\left(2\zeta-1 \right) \sin(\phi) \sin(\theta) \rho_{a'} \rho^{(0)} \rho_{a},
\end{eqnarray}
where we have taken into account that $\rho_{a'} \rho^{(0)} \rho_{a} = \rho_{a} \rho^{(0)} \rho_{a'}$ because of the diagonality of each factor. If the control qubit is ignored, an incoherent mixture of the two causal orders of $\mathbf{A}$ and $\mathbf{A}'$, due to the probability distribution given by the diagonal elements of $\omega$, is realized. Due to Eq.~\eqref{eq:incStrength}, this implements an effective measurement strength of
\begin{eqnarray}\label{app:incStrengthUnc}
    \nonumber \bar{a}^{\rm inc}_\text{unc}(\theta, \zeta) &=& \frac{1}{2}\Bigl(1 + (2\zeta-1)\cos(\theta)\Bigr)a' \\
    &+& \frac{1}{2}\Bigl(1 - (2\zeta-1)\cos(\theta)\Bigr) a.
\end{eqnarray}
On the other hand, $\Lambda_X^{(a,a')}$ and $\Lambda_Y^{(a,a')}$ and, in turn, the corresponding contribution of SCO
\begin{equation}
     \Xi_{\phi'}^{(a,a')}[\varrho_{\rm in}^{\rm{unc}}] =  (2\zeta - 1) \sin(\theta) \cos(\phi-\phi') \rho_{a'} \rho^{(0)} \rho_{a}
\end{equation}
carry no coherence. Therefore, a measurement of $\mathbf{C}$ is equivalent to the implementation of one of the effective measurement channels of strengths \eqref{app:eq:postSelStrength}
with
\begin{equation}
    \Delta^{(a,a')}_{\rm unc} = \frac{1}{2}aa'(2\zeta - 1) \sin(\theta) \cos(\phi - \phi') (1 + \tanh (\beta \varepsilon)),
\end{equation}
resulting in
\begin{eqnarray}
    \nonumber W_{\rm SCO}^{\rm diag}(\varrho_{\rm in}^{\rm unc}) &=& -(2\zeta - 1) \delta_{a,a'}(\beta\varepsilon)\sin(\theta) \cos(\phi - \phi'), \\
    W_{\rm SCO}^{\rm off}(\varrho_{\rm in}^{\rm unc}) &=& 0,
\end{eqnarray}
where
\begin{eqnarray}
    \nonumber \delta_{a,a'}(\beta\varepsilon) &=& \frac{1}{2}\Bigl((1-a)(1-a') (1 - \tanh (\beta \varepsilon)) \\
    &-& aa' (1 + \tanh (\beta \varepsilon))\Bigr).
\end{eqnarray}
Hence, applying \eqref{eq:nonZeroEffCondGen}, we obtain that SCO produces a non-zero contribution to the efficiency of the heat engine if
\begin{equation}
    |1-2\bar{a}^{\rm inc}_{\rm unc}(\theta, \zeta)| < \Bigl| (2\zeta - 1)\delta_{a,a'}(\beta\varepsilon) \sin(\theta) \cos(\phi - \phi')\Bigr|,
\end{equation}
implying that the maximal efficiency \eqref{eq:etaSwitch} is achieved for initial states with $\zeta \in \{0,1\}$ (i.e., a pure $\omega$) and $\phi \in \{\phi' + \pi k \mid k \in \mathbb{Z}\}$:
\begin{equation}\label{eq:uncEtaOpt}
    \max_{\zeta, \phi} \tilde{\eta}^{\rm coh}_{\rm unc} = \frac{| \delta_{a,a'}(\beta\varepsilon)\sin(\theta)| + 1-2\bar{a}^{\rm inc}_{\rm unc}(\theta)}{1-2\bar{a}^{\rm inc}_{\rm unc}(\theta)+\tanh(\beta\varepsilon)},
\end{equation}
where
\begin{equation}
    \bar{a}^{\rm inc}_{\rm unc}(\theta) = \frac{1}{2}\Bigl(a + a' + |a-a'||\cos(\theta)|\Bigr).
\end{equation}
Further optimization over $\theta$ depends sensitively on the measurement strengths $a$ and $a'$ of the apparata arranged in indefinite causal order, as well as on the initial temperature of the working medium $\mathbf{Q}$. In particular, for the case of complementary apparata (i.e., $a' = 1-a$), the efficiency $\tilde{\eta}^{\rm coh}_{\rm unc}$ exhibits three distinct parameter regions in strength--temperature space (see Fig.~\ref{fig:optangle_unc}). First, below the line
\begin{equation}\label{eq:borderHeatCond}
|1-2a| = \tanh(\beta\varepsilon),
\end{equation}
the constraints \eqref{eq:heatEngCondSwitch1}--\eqref{eq:heatEngCondSwitch2} are not necessarily fulfilled for all values of $\theta \in [0,\pi]$, so the optimization must be restricted to a reduced interval of angles. This restriction is particularly relevant at high temperatures (i.e., for small $\beta\varepsilon$). Conversely, above this line, two regimes can be distinguished: one in which the optimal angles cluster near $\theta_{\rm opt} \simeq 0$, and another in which they cluster near $\theta_{\rm opt} \simeq \pi/2$. The location of the optimum is determined by whether the first (arising from SCO) or the second term in \eqref{eq:uncEtaOpt} dominates.

Similarly to $\tilde{\eta}^{\rm coh}$, the efficiencies $\eta^{(\pm)}$ associated with measurement outcomes of $\mathbf{C}$ are maximized for a pure initial state $\omega$, so that $\zeta \in \{0, 1\}$, and $\phi \in \{\phi' + \pi k \mid k \in \mathbb{Z}\}$,
\begin{widetext}
\begin{equation}\label{eq:uncEtaPostSelOpt}
    \max_{\zeta, \phi} \eta^{(\pm)}_{\rm unc} = \frac{\Bigl| 1-2\bar{a}^{\rm inc}_{\rm unc}(\theta) + \delta_{a,a'}(\beta\varepsilon) \sin(\theta)\Bigr| + 1-2\bar{a}^{\rm inc}_{\rm unc}(\theta) + \delta_{a,a'}(\beta\varepsilon)\sin(\theta)}{1-2\bar{a}^{\rm inc}_{\rm unc}(\theta) + \delta_{a,a'}(\beta\varepsilon)\sin(\theta) + \Bigl(1 - \Sigma_{a,a'}(\beta\varepsilon)\sin(\theta)\Bigr)\tanh(\beta\varepsilon)},
\end{equation}
\end{widetext}
where
\begin{eqnarray}
    \nonumber \Sigma_{a,a'}(\beta\varepsilon) &=& \frac{1}{2}\Bigl((1-a)(1-a') (1 - \tanh (\beta \varepsilon)) \\
    &+& aa' (1 + \tanh (\beta \varepsilon))\Bigr).
\end{eqnarray}

\subsection{Correlated initial state} \label{app:cc_purity}

For scenarios where the working medium $\mathbf{Q}$ and the controller $\mathbf{C}$ share a priori correlations, we consider the initial joint state
\begin{eqnarray} \label{app:rho_ini_entangled}
        \nonumber\varrho^{\rm qe}_{\rm in}&=& \sum_{\ell = 0,1}\frac{1+{(-1)}^\ell \tanh(\beta\varepsilon)}{2} \ketbra{\ell} \otimes \omega_\ell \\ &+& \xi_{\beta\varepsilon} e^{i\varphi}\ket{0} \bra{1} \otimes \ket{\psi_{\theta,\phi}}_{\rm c}\bra{\psi_{\theta-\pi,\phi}} +\text{h.c.},
\end{eqnarray}
where
\begin{equation}
    \omega_\ell = \zeta_\ell \ketbrac{\psi_{\theta,\phi}}+ (1-\zeta_\ell) \ketbrac{\psi_{\theta-\pi,\phi}},
\end{equation}
with the state $\ket{\psi_{\theta,\phi}}$ given by \eqref{app:pureState}. The parameters $\xi_{\beta\varepsilon} \in \Bigl[0, \frac{\operatorname{sech}(\beta\varepsilon)}{2}\Bigr]$ and $\zeta_{0,1} \in [0,1 ]$ characterize the strength of correlations shared between $\mathbf{Q}$ and $\mathbf{C}$, and $\varphi \in [0, 2\pi]$.

\begin{figure}[t!]
    \centering
    \includegraphics[width=0.9\columnwidth]{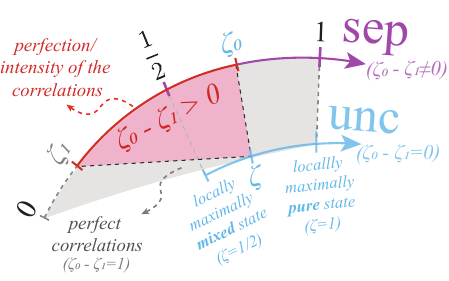}
    \caption[Initial correlations parameter space for initial states that allow a separable representation]{\textbf{Initial correlations parameter space for initial states that allow a separable representation.} The purple curve represents all admissible $(\zeta_0,\zeta_1)$ satisfying $0 \leq \zeta_1 \leq \zeta_0$ and $1/2 \leq \zeta_0 \leq 1$. The red segment highlights the degree of correlation $\zeta_0-\zeta_1$. Perfect correlations ($\zeta_0-\zeta_1=1$) occur when the pink and gray regions coincide, while the uncorrelated limit ($\zeta_0=\zeta_1$) corresponds to the blue curve, interpolating between locally maximally mixed and maximally pure control states. Increasing either $\zeta$ (in the uncorrelated case) or $\zeta_0-\zeta_1$ (at fixed $\zeta_0$) enhances the SCO effect.}
    \label{fig:parametros}
\end{figure}

\emph{Separable joint initial state.} When $\xi_{\beta\varepsilon} = 0$, \eqref{app:rho_ini_entangled} reduces to the separable state
\begin{eqnarray}\label{app:rho_cl}
            \nonumber \varrho^{\rm sep}_{\textrm{in}} &:=& \sum_{\ell = 0,1}\frac{1+{(-1)}^\ell \tanh(\beta\varepsilon)}{2} \ketbra{\ell} \otimes \omega_\ell,
\end{eqnarray}
which exhibits correlations of classical nature governed by the difference $\zeta_0-\zeta_1$. Without loss of generality, we impose $0 \leq \zeta_1 \leq \zeta_0 \leq 1$ and $\zeta_0 \geq \frac{1}{2}$, so that $\zeta_0 -\zeta_1 = 0$ reproduces the uncorrelated case, and $\zeta_0 - \zeta_1 = 1$ corresponds to perfect separable correlations (see Fig.~\ref{fig:parametros}). Application of the measurement apparata $\mathbf{A}$ and $\mathbf{A}'$ via the quantum SWITCH produces the state \eqref{app:SwitchGeneralMeasChan} with
\begin{eqnarray}\label{app:SwitchGeneralMeasChanSep}
    \lambda[\varrho^{\rm sep}_{\textrm{in}}] &=& \frac{1}{2}\Bigl(1 + (2\operatorname{tr}[\mathfrak{Z}\rho^{(0)}]-1)\cos(\theta)\Bigr), \\
    \Lambda_X^{(a,a')}[\varrho^{\rm sep}_{\textrm{in}}] &=& \frac{1}{2}\cos(\phi) \sin(\theta) \rho_{a'} (2\mathfrak{Z} - \mathbb{I}) \rho^{(0)} \rho_{a}, \\
    \Lambda_Y^{(a,a')}[\varrho^{\rm sep}_{\textrm{in}}] &=& \frac{1}{2}\sin(\phi) \sin(\theta) \rho_{a'} (2\mathfrak{Z} - \mathbb{I} ) \rho^{(0)} \rho_{a},
\end{eqnarray}
where we have taken into account that $\rho_{a'} \rho^{(0)} \rho_{a} = \rho_{a} \rho^{(0)} \rho_{a'}$ because of the diagonality of each factor. In the incoherent regime of the quantum SWITCH, when the control qubit $\mathbf{C}$ is not observed, an effective measurement channel of strength
\begin{eqnarray}\label{app:incStrengthSep}
    \nonumber \bar{a}^{\rm inc}_\text{sep}(\theta, \zeta_{0,1}) &=& \frac{1}{2}\Bigl(1 + (2\operatorname{tr}[\mathfrak{Z}\rho^{(0)}]-1)\cos(\theta)\Bigr)a' \\
    &+& \frac{1}{2}\Bigl(1 - (2\operatorname{tr}[\mathfrak{Z}\rho^{(0)}]-1)\cos(\theta)\Bigr) a,
\end{eqnarray}
is applied to the working medium $\mathbf{Q}$, where $\mathfrak{Z} = \text{diag}(\zeta_0, \zeta_1)$.

\begin{figure}[t!]
    \centering
    \includegraphics[width=\columnwidth]{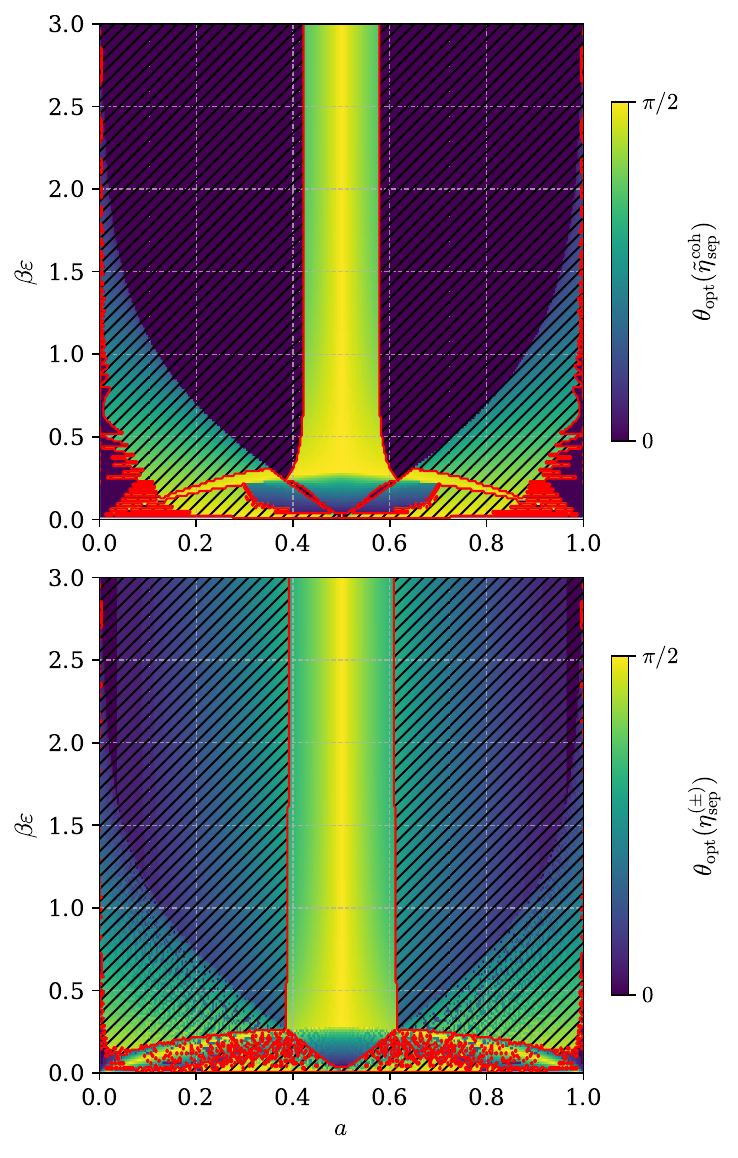}
    \caption[Optimal angles $\theta$ maximizing the efficiencies $\tilde{\eta}^{\rm coh}$ and $\eta^{(\pm)}$ for a separable initial state.]{\textbf{Optimal angles $\theta$ maximizing the efficiencies $\tilde{\eta}^{\rm coh}$ and $\eta^{(\pm)}$ for a separable initial state.} The plots show the optimal measurement angle $\theta_{\rm opt}$ that maximizes the efficiency of the heat engine as a function of the measurement strength and the inverse temperature, for a separable state $\varrho_{\rm in}^{\rm sep}$ and measurement apparata of complementary strengths $a$ and $a' = 1-a$. The upper panel presents $\theta_{\rm opt}$ corresponding to the efficiency $\tilde{\eta}^{\rm coh}$, while the lower panel presents $\theta_{\rm opt}$ for the efficiencies $\eta^{(\pm)}$. The hatched regions indicate parameter configurations in which the separable state $\varrho_{\rm in}^{\rm sep}$ yields no efficiency advantage over the uncorrelated state $\varrho_{\rm in}^{\rm unc}$ because of violation of heat engine conditions.}
    \label{fig:optangle_sep}
\end{figure}

In contrast to \eqref{app:incStrengthUnc}, even in the presence of purely classical correlations, the action of the quantum SWITCH in the incoherent regime becomes dependent on the initial temperature of $\mathbf{Q}$ through the purity of the local state of $\mathbf{C}$, as determined by
\begin{eqnarray}\label{app:localStateSep}
        \nonumber \operatorname{tr}_\text{q}[\varrho_{\rm in}^{\rm sep}] &=& \operatorname{tr}[\mathfrak{Z}\rho^{(0)}] \ketbrac{\psi_{\theta,\phi}} \\
        &+& \Bigl(1 - \operatorname{tr}[\mathfrak{Z}\rho^{(0)}]\Bigr) \ketbrac{\psi_{\theta-\pi,\phi}}.
\end{eqnarray}
On the other hand, similarly to the case of initially uncorrelated $\mathbf{Q}$ and $\mathbf{C}$, no coherence is carried by $\Lambda_X^{(a,a')}$ and $\Lambda_Y^{(a,a')}$ and, in turn, the corresponding contribution of SCO
\begin{equation}
     \Xi_{\phi'}^{(a,a')}[\varrho_{\rm in}^{\rm{sep}}] =  \sin(\theta) \cos(\phi-\phi') \rho_{a'} (2\mathfrak{Z} - \mathbb{I}) \rho^{(0)} \rho_{a}
\end{equation}
carries no coherence. Therefore, a measurement of $\mathbf{C}$ is also equivalent to the implementation of one of the effective measurement channels of strengths \eqref{app:eq:postSelStrength} with
\begin{equation} \label{eq:a_bar_cl}
    \Delta^{(a,a')}_{\rm sep} = \frac{1}{2}aa'(2\zeta_0 - 1) \sin(\theta) \cos(\phi-\phi') (1 + \tanh (\beta \varepsilon)),
\end{equation}
resulting in
\begin{eqnarray}
    \nonumber W_{\rm SCO}^{\rm diag}(\varrho_{\rm in}^{\rm sep}) &=& \sin(\theta) \cos(\phi - \phi')\Bigl( (\zeta_0 - \zeta_1) \Sigma_{a,a'}(\beta\varepsilon) \\
    &-& (\zeta_0 + \zeta_1 - 1) \delta_{a,a'}(\beta\varepsilon) \Bigr) , \\
    W_{\rm SCO}^{\rm off}(\varrho_{\rm in}^{\rm sep}) &=& 0.
\end{eqnarray}
Applying \eqref{eq:nonZeroEffCond}, SCO yields a non-zero contribution to the heat engine efficiency whenever
\begin{align}
    |1&-2\bar{a}^{\rm inc}_{\rm sep}(\theta, \zeta_{0,1})| < \Bigl| \Bigl( (\zeta_0 - \zeta_1) \Sigma_{a,a'}(\beta\varepsilon) \nonumber \\
    & - (\zeta_0 + \zeta_1 - 1)\delta_{a,a'}(\beta\varepsilon) \Bigr)  \sin(\theta) \cos(\phi - \phi')\Bigr|,
\end{align}
so that it is maximized by $\phi \in \{\phi' + \pi k \mid k\in\mathbb{Z}\}$ and $\zeta_{0,1} \in \{0, 1\}$. These limiting values of $\zeta_{0,1}$ correspond to two distinct configurations: (i) $\zeta_0=\zeta_1$ yields the uncorrelated initial state $\varrho_{\rm in}^{\rm unc}$, while (ii) $\zeta_0=1$, $\zeta_1=0$ represents a perfectly separable correlated state $\varrho_{\rm in}^{\rm sep}$. Since $\Sigma_{a,a'}(\beta\varepsilon)\geq\delta_{a,a'}(\beta\varepsilon)$, the separable configuration generally provides a stronger SCO contribution and thus a higher potential efficiency, provided that the heat engine conditions \eqref{app:eq:aLowBoundSwitch}–\eqref{app:eq:aUppBoundSwitch} are not violated under both measurement outcomes of $\mathbf{C}$. In that case, the optimized efficiency reads
\begin{equation}\label{eq:eta_sep_preopt}
    \max_{\zeta_{0}-\zeta_{1}=1, \phi}\tilde{\eta}^{\rm coh}_{\rm sep} = \frac{| \Sigma_{a,a'}(\beta\varepsilon)\sin(\theta)| + 1-2\bar{a}^{\rm inc}_{\rm sep}(\theta)}{1-2\bar{a}^{\rm inc}_{\rm sep}(\theta)+\tanh(\beta\varepsilon)},
\end{equation}
where
\begin{equation}
    \bar{a}^{\rm inc}_{\rm sep}(\theta) = \frac{1}{2}\Bigl(a + a' + |a-a'||\cos(\theta)|\tanh(\beta\varepsilon)\Bigr).
\end{equation}
Conversely, if the separable configuration is incompatible with the conditions \eqref{app:eq:aLowBoundSwitch}–\eqref{app:eq:aUppBoundSwitch}, the optimization reverts to the uncorrelated case, where the best achievable efficiency is given by \eqref{eq:uncEtaOpt}. Similarly to the uncorrelated scenario, the subsequent optimization over $\theta$ depends sensitively on them as well as on $\beta\varepsilon$. In particular, for complementary apparata ($a' = 1 - a$), the perfectly correlated separable state $\varrho^{\rm sep,\,\zeta_0 - \zeta_1 = 1}_{\rm in}$ provides an advantage mainly in the vicinity of $a = 1/2$. For other values of the measurement strength, the optimization instead favors the uncorrelated configuration once the initial temperature of $\mathbf{Q}$ decreases (i.e., when $\beta\varepsilon$ grows). As a result, three distinct regions can again be identified in the strength–temperature plane (see Fig.~\ref{fig:optangle_sep}). Relative to the uncorrelated case, the separable correlated state expands the parameter region in which the SCO-induced term in \eqref{eq:eta_sep_preopt} dominates the efficiency, while below the line \eqref{eq:borderHeatCond} the set of angles satisfying the heat engine conditions \eqref{app:eq:aLowBoundSwitch}–\eqref{app:eq:aUppBoundSwitch} is correspondingly reduced.

Similarly to $\tilde{\eta}^{\rm coh}$, the efficiencies $\eta^{(\pm)}$ associated with measurement outcomes of $\mathbf{C}$ are maximized for $\zeta \in \{0, 1\}$ and $\phi \in \{\phi' + \pi k \mid k \in \mathbb{Z}\}$. If the corresponding heat engine conditions \eqref{app:eq:aLowBoundSwitch}–\eqref{app:eq:aUppBoundSwitch} are satisfied under the respective measurement outcome of $\mathbf{C}$ for $\bar{a}_+$ or $\bar{a}_-$, the optimized efficiency reads
\begin{widetext}
\begin{equation}\label{eq:uncEtaPostSelOptSep}
    \max_{\zeta_0-\zeta_1=1, \phi} \eta^{(\pm)}_{\rm sep} = \frac{\Bigl| 1-2\bar{a}^{\rm inc}_{\rm sep}(\theta) + \Sigma_{a,a'}(\beta\varepsilon) \sin(\theta)\Bigr| + 1-2\bar{a}^{\rm inc}_{\rm sep}(\theta) + \Sigma_{a,a'}(\beta\varepsilon)\sin(\theta)}{1-2\bar{a}^{\rm inc}_{\rm sep}(\theta) + \Sigma_{a,a'}(\beta\varepsilon)\sin(\theta) + \Bigl(1 - \delta_{a,a'}(\beta\varepsilon)\sin(\theta)\Bigr)\tanh(\beta\varepsilon)}.
\end{equation}
\end{widetext}
If these conditions are violated, the optimal efficiency reduces to that of the uncorrelated case given in \eqref{eq:uncEtaPostSelOpt}.

\begin{figure}[t!]
    \centering
    \includegraphics[width=\columnwidth]{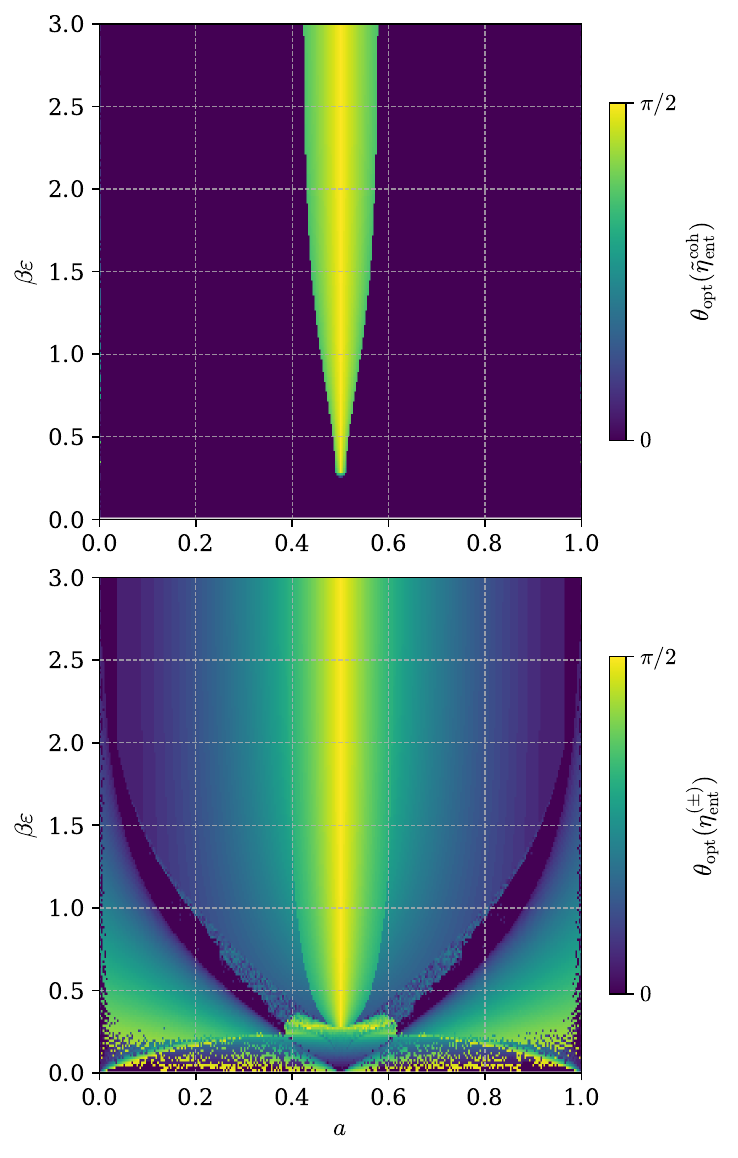}
    \caption[Optimal angles $\theta$ maximizing the efficiencies $\tilde{\eta}^{\rm coh}$ and $\eta^{(\pm)}$ for an uncorrelated initial state.]{\textbf{Optimal angles $\theta$ maximizing the efficiencies $\tilde{\eta}^{\rm coh}$ and $\eta^{(\pm)}$ for an entangled initial state.} The plots show the optimal measurement angle $\theta_{\rm opt}$ that maximizes the efficiency of the heat engine as a function of the measurement strength and the inverse temperature, for an entangled initial state $\varrho_{\rm in}^{\rm qe}$ and measurement apparata of complementary strengths $a$ and $a' = 1-a$. The upper panel presents $\theta_{\rm opt}$ corresponding to the efficiency $\tilde{\eta}^{\rm coh}$, while the lower panel presents $\theta_{\rm opt}$ for the efficiencies $\eta^{(\pm)}$.}
    \label{fig:optangle_ent}
\end{figure}

\emph{Entangled joint initial state.} Allowing for non-zero $\xi_{\beta\varepsilon}$, we proceed with entangled initial states \eqref{app:rho_ini_entangled}, for which the quantum SWITCH produces the state \eqref{app:SwitchGeneralMeasChan} with
\begin{eqnarray*}
    \Lambda_X^{(a,a')}[\varrho^{\rm qe}_{\rm in}] &=& \Lambda_X^{(a,a')}[\varrho^{\rm sep}_{\textrm{in}}] + \frac{\xi_{\beta\varepsilon}}{2}\Bigl[e^{-i \phi} \mathfrak{E}_{\theta, \varphi}^{(a,a')} + \text{h.c.}\Bigr], \\
    \Lambda_Y^{(a,a')}[\varrho^{\rm qe}_{\rm in}] &=& \Lambda_Y^{(a,a')}[\varrho^{\rm sep}_{\textrm{in}}] + \frac{i\xi_{\beta\varepsilon}}{2}\Bigl[e^{-i \phi}\mathfrak{E}_{\theta, \varphi}^{(a,a')} - \text{h.c.}\Bigr],
\end{eqnarray*}
where
\begin{align}
    \mathfrak{E}_{\theta, \varphi}^{(a,a')} &= \rho_{a'} \Bigl(- \cos^2 \Bigl(\frac{\theta}{2} \Bigr)e^{-i\varphi}\ket{0}\bra{1} \nonumber \\
    &\qquad + \sin^2 \Bigl(\frac{\theta}{2} \Bigr)e^{i\varphi}\ket{1}\bra{0}\Bigr)\rho_a \\
    &= e^{-i {\varphi Z}/{2}} \rho_{a'} \Bigl(- \cos^2 \Bigl(\frac{\theta}{2} \Bigr)\ket{0}\bra{1} \nonumber \\
    &\qquad + \sin^2 \Bigl(\frac{\theta}{2} \Bigr)\ket{1}\bra{0} \Bigr)\rho_a e^{i {\varphi Z}/{2}} \label{eq:offDiagXi} \\
    &:= e^{-i {\varphi Z}/{2}} \mathfrak{E}_{\theta}^{(a,a')} e^{i {\varphi Z}/{2}}.
\end{align}
Therefore, the contribution of SCO takes the form
\begin{align}
     \Xi_{\phi'}^{(a,a')}[\varrho_{\rm in}^{\rm qe}] &= \Xi_{\phi'}^{(a,a')}[\varrho_{\rm in}^{\rm{sep}}] + \xi_{\beta\varepsilon} \tilde{\mathfrak{E}}_{\theta, \phi-\phi', \varphi}^{(a,a')}, \label{eq:XiEnt}
\end{align}
where we denote $\tilde{\mathfrak{E}}_{\theta, \phi-\phi', \varphi}^{(a,a')} = e^{-i (\phi-\phi')} \mathfrak{E}_{\theta, \varphi}^{(a,a')} + \text{h.c.}$, which may carry coherence. In the presence of initial entanglement between $\mathbf{Q}$ and $\mathbf{C}$, the coherent regime of the quantum SWITCH can thus contribute to both components of $\vec{W}_{\rm SCO}$. Specifically, the first term in \eqref{eq:XiEnt} determines
\begin{equation}
    W_{\rm SCO}^{\rm diag}(\varrho_{\rm in}^{\rm qe}) = W_{\rm SCO}^{\rm diag}(\varrho_{\rm in}^{\rm sep}),
\end{equation}
which coincides with the corresponding value obtained for a separable initial state. By contrast, the second term affects only the off-diagonal elements of the state and therefore contributes exclusively to $W_{\rm SCO}^{\rm off}(\varrho_{\rm in}^{\rm qe})$. To analyze this contribution, we consider the Hilbert--Schmidt norm of the second term in \eqref{eq:XiEnt}:
\begin{eqnarray}
    \nonumber \xi_{\beta\varepsilon}\lVert \tilde{\mathfrak{E}}_{\theta, \phi-\phi', \varphi}^{(a,a')} \rVert_{\rm HS} &=& \xi_{\beta\varepsilon} \sqrt{\operatorname{tr}[(\tilde{\mathfrak{E}}_{\theta, \phi-\phi', \varphi}^{(a,a')})^2]} \\
    \nonumber &=& \xi_{\beta\varepsilon} \sqrt{\sum_\ell \langle\ell|(\tilde{\mathfrak{E}}_{\theta, \phi-\phi', \varphi}^{(a,a')})^2|\ell\rangle} \\
    \nonumber &=& \xi_{\beta\varepsilon} \sqrt{\sum_{\ell,\ell'} \langle\ell|\tilde{\mathfrak{E}}_{\theta, \phi-\phi', \varphi}^{(a,a')}|\ell'\rangle \langle\ell'|\tilde{\mathfrak{E}}_{\theta, \phi-\phi', \varphi}^{(a,a')}|\ell\rangle} \\
    \nonumber &=& \xi_{\beta\varepsilon}\sqrt{\sum_{\ell \neq \ell'} \langle\ell|\tilde{\mathfrak{E}}_{\theta, \phi-\phi', \varphi}^{(a,a')}|\ell'\rangle \langle\ell'|\tilde{\mathfrak{E}}_{\theta, \phi-\phi', \varphi}^{(a,a')}|\ell\rangle} \\
    \nonumber &=& \xi_{\beta\varepsilon} \sqrt{2} \Bigl|\operatorname{tr}\Bigl[\tilde{\mathfrak{E}}_{\theta, \phi-\phi', \varphi}^{(a,a')}|1\rangle\langle 0|\Bigr]\Bigr| \\
    \nonumber &=& \sqrt{2} \Bigl|\operatorname{tr}\Bigl[\Xi_{\phi'}^{(a,a')}[\varrho_{\rm in}^{\rm qe}]|1\rangle\langle 0|\Bigr]\Bigr| \\
    &=& \frac{1}{\sqrt{2}}W_{\rm SCO}^{\rm off}(\varrho_{\rm in}^{\rm qe}),
\end{eqnarray}
where we use the definition $\lVert X \rVert_{\rm HS} = \sqrt{\operatorname{tr}[X^\dagger X]}$ of the Hilbert--Schmidt norm for any suitable operator $X$ on the Hilbert space of $\mathbf{Q}$ and take into account that the diagonal elements of $\mathfrak{E}_{\theta, \varphi}^{(a,a')}$ vanish. On the other hand, calculating the Hilbert--Schmidt norm explicitly, we have
\begin{align}
    \lVert \tilde{\mathfrak{E}}_{\theta, \phi-\phi', \varphi}^{(a,a')} \rVert_{\rm HS}^2 &= \operatorname{tr}[(\tilde{\mathfrak{E}}_{\theta, \phi-\phi', \varphi}^{(a,a')})^2] \\
    &= \operatorname{tr}[\mathfrak{E}_{\theta, \varphi}^{(a,a')} \mathfrak{E}_{\theta, \varphi}^{(a,a')\dagger} + \mathfrak{E}_{\theta, \varphi}^{(a,a')\dagger} \mathfrak{E}_{\theta, \varphi}^{(a,a')} \nonumber \\
    &\qquad + 2\cos(2 (\phi-\phi')) (\mathfrak{E}_{\theta, \varphi}^{(a,a')})^2] \\
    &= \operatorname{tr}[(\mathfrak{E}_{\theta, \varphi}^{(a,a')} + \mathfrak{E}_{\theta, \varphi}^{(a,a')\dagger})^2 \nonumber \\
    &\qquad - 4\sin^2(\phi-\phi') (\mathfrak{E}_{\theta, \varphi}^{(a,a')})^2] \\
    &= \lVert \mathfrak{E}_{\theta, \varphi}^{(a,a')} + \mathfrak{E}_{\theta, \varphi}^{(a,a')\dagger}\rVert_{\rm HS}^2 \nonumber \\
    &\qquad - 4\sin^2(\phi-\phi') \operatorname{tr}[(\mathfrak{E}_{\theta, \varphi}^{(a,a')})^2].
\end{align}
Therefore, we obtain
\begin{widetext}
\begin{eqnarray}\label{eq:cohWorkEnt}
W_{\rm SCO}^{\rm off}(\varrho_{\rm in}^{\rm qe}) &=&  \xi_{\beta\varepsilon} \sqrt{2} \sqrt{\lVert \mathfrak{E}_{\theta, \varphi}^{(a,a')} + \mathfrak{E}_{\theta, \varphi}^{(a,a')\dagger}\rVert_{\rm HS}^2 - 4\sin^2(\phi-\phi') \operatorname{tr}[(\mathfrak{E}_{\theta, \varphi}^{(a,a')})^2]} \\
    &=& 2\xi_{\beta\varepsilon} \sqrt{\Bigl|a(1-a')\sin^2\Bigl(\frac{\theta}{2}\Bigr) - a'(1-a)\cos^2\Bigl(\frac{\theta}{2}\Bigr)\Bigr|^2 + aa'(1-a)(1-a')\sin^2(\theta) \sin^2(\phi-\phi')}, \nonumber
\end{eqnarray}
\end{widetext}
which is invariant with respect to $\varphi$. Optimization over the parameters of the initial state $\varrho_{\rm in}^{\rm qe}$ shows that the heat engine efficiency $\tilde{\eta}^{\rm coh}$ reaches its maximum for $\zeta_{0,1} \in \{0,1\}$ and the largest admissible $\xi_{\beta\varepsilon} >0$ consistent with the heat engine constraints \eqref{app:eq:aLowBoundSwitch}–\eqref{app:eq:aUppBoundSwitch} under both measurement outcomes of $\mathbf{C}$. The latter becomes particularly relevant at high initial temperatures of $\mathbf{Q}$ (i.e., low $\beta\varepsilon$). Contrary to the uncorrelated and separable cases ($\varrho^{\rm unc}_{\rm in}$ and $\varrho^{\rm sep}_{\rm in}$), the optimization over the phase $\phi$ for the entangled initial state yields two distinct phase values that maximize the SCO contribution, $\phi \in \{\phi' + \pi k \mid k \in \mathbb{Z}\}$ and $\phi \in \{\phi' + \frac{\pi}{2} + \pi k \mid k \in \mathbb{Z}\}$. Which of these attains the global maximum depends on the dominant component of $\vec{W}_{\rm SCO}(\varrho_{\rm in}^{\rm qe})$: the first choice enhances $|W_{\rm SCO}^{\rm diag}(\varrho_{\rm in}^{\rm qe})|$, reflecting population imbalance in the working medium, whereas the second one maximizes $|W_{\rm SCO}^{\rm off}(\varrho_{\rm in}^{\rm qe})|$, related to the consumption of state coherence. Consequently,
\begin{equation}
    \max_{\phi}\tilde{\eta}^{\rm coh}  = \frac{\max[\langle \mathcal{W}_{\rm ext}^{\rm coh} \rangle_{\phi'}, \langle \mathcal{W}_{\rm ext}^{\rm coh} \rangle_{\phi' + \pi/2}]}{1-2\bar{a}^{\rm inc}+\tanh(\beta\varepsilon)},
\end{equation}
where
\begin{widetext}
\begin{eqnarray}
    \langle \mathcal{W}_{\rm ext}^{\rm coh} \rangle_{\phi'} &=& \frac{1}{2}\Bigl(\sqrt{(1-2\bar{a}^{\rm inc} + W_{\rm \phi'}^{\rm diag})^2 + |W_{\rm \phi'}^{\rm off}|^2} + \sqrt{(1-2\bar{a}^{\rm inc} - W_{\rm \phi'}^{\rm diag})^2 + |W_{\rm \phi'}^{\rm off}|^2}\Bigr) + 1-2\bar{a}^{\rm inc}, \\
    \langle \mathcal{W}_{\rm ext}^{\rm coh} \rangle_{\phi' + \pi/2} &=& \sqrt{(1-2\bar{a}^{\rm inc})^2 + 4\xi_{\beta\varepsilon}^2 \Bigl(a(1-a')\sin^2\Bigl(\frac{\theta}{2}\Bigr) + a'(1-a)\cos^2\Bigl(\frac{\theta}{2}\Bigr)\Bigr)^2} + 1-2\bar{a}^{\rm inc},
\end{eqnarray}
\end{widetext}
with
\begin{align}
    W_{\rm \phi'}^{\rm diag} &= \sin(\theta)\Bigl((\zeta_0 - \zeta_1) \Sigma_{a,a'}(\beta\varepsilon) \nonumber \\
    &\qquad - (\zeta_0 + \zeta_1 - 1) \delta_{a,a'}(\beta\varepsilon) \Bigr), \label{app:eq:diagEntOptWork} \\
    W_{\rm \phi'}^{\rm off} &= 2\xi_{\beta\varepsilon} \Bigl|a(1-a')\sin^2\Bigl(\frac{\theta}{2}\Bigr) - a'(1-a)\cos^2\Bigl(\frac{\theta}{2}\Bigr)\Bigr|. \label{app:eq:offEntOptWork}
\end{align}
Further optimization of $\tilde{\eta}^{\rm coh}$ depends on the measurement strengths $a$ and $a'$ as well as on the initial temperature of the working medium. The admissible range of the correlation parameter, $\xi_{\beta\varepsilon} \in \bigl[0,\frac{\operatorname{sech}(\beta\varepsilon)}{2}\bigr]$, decreases with increasing $\beta\varepsilon$, implying that the contribution of the off-diagonal component $|W_{\rm SCO}^{\rm off}(\varrho_{\rm in}^{\rm qe})|$ becomes dominant at high initial temperatures (i.e., for small $\beta\varepsilon$) of $\mathbf{Q}$.

\begin{figure*}[t!]
    \centering
    \includegraphics[width=\linewidth]{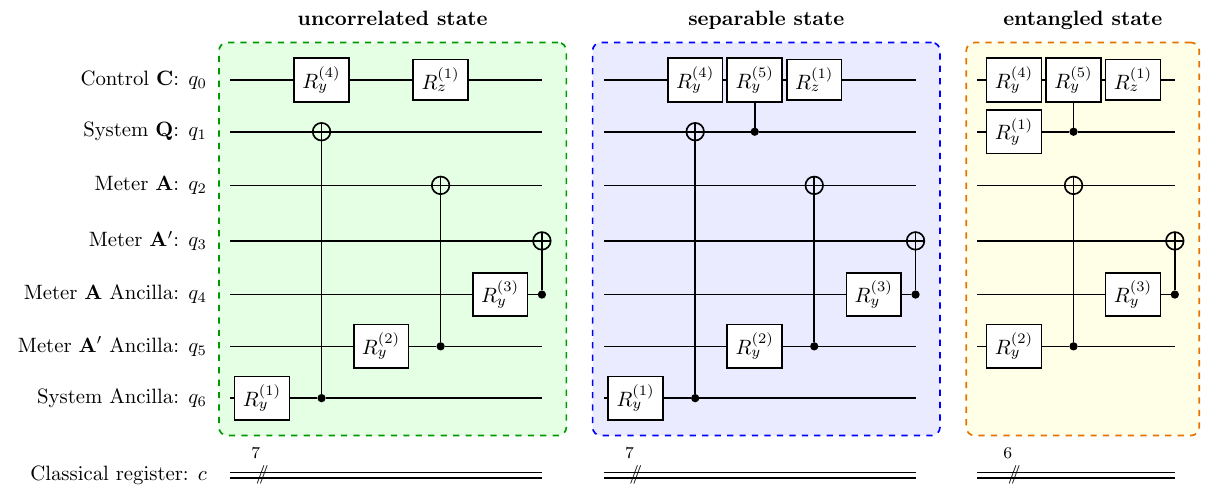}
    \caption[Preparation of the initial states for the three correlation regimes.]{\textbf{Preparation of the initial states for the three correlation regimes.} Qubits $q_0$ and $q_1$ represent the control~$\mathbf{C}$ and working medium~$\mathbf{Q}$, respectively. Qubits $q_2$-$q_3$ act as \emph{meter ancillae}, implementing the measurement operations $\mathcal{M}^a$ and $\mathcal{M}^{a'}$ within the quantum SWITCH and mediating the effective heat exchange between the meters and the working medium. Qubits $q_4$-$q_6$ (or $q_4$ and $q_5$ in the case of the entangled initial state) serve as \emph{auxiliary ancillae} for the preparation of the desired initial states of $q_0$-$q_3$, while $c$ denotes the corresponding classical registers. The single-qubit rotations $R_y^{(1-4)}(\theta_y^{(1-4)})$ around the $y$-axis and $R_z^{(1)}(\theta_z^{(1)})$ around the $z$-axis are determined by the measurement strengths $a$ and $a' = 1-a$, together with the Gibbs-state parameter $\beta\varepsilon$. The corresponding rotation angles are $\theta_y^{(1)} = \arccos{(\sqrt{\tanh(\beta\varepsilon)})}$, $\theta_y^{(2)} = \arccos{(\sqrt{2a-1})}$, $\theta_y^{(3)} = \arccos{(\sqrt{2a'-1})}$, $\theta_y^{(4)} = \pi/2$, and $\theta_z^{(1)} = \phi/4$. The controlled rotation $R_y^{(5)}(-\pi)$ establishes the initial correlations between $\mathbf{Q}$ and $\mathbf{C}$. After the preparation stage, qubits $q_4$-$q_6$ (or $q_4$ and $q_5$ for the entangled initial state) are traced out.}
    \label{fig:preparation}
\end{figure*}

For complementary measurement apparata ($a' = 1-a$), the presence of initial quantum entanglement primarily affects the region below the line \eqref{eq:borderHeatCond}. In most of the parameter space $(a, \beta\varepsilon)$, the optimal efficiency stems from state coherence, since $W_{\phi'}^{\rm diag}=0$ when $\theta=0$. In contrast, the advantage due to population imbalance becomes relevant mainly in the vicinity of $a = \tfrac{1}{2}$, where the optimal angles cluster around $\theta_{\rm opt} \simeq \pi/2$ (see Fig.~\ref{fig:optangle_ent}). The optimization of the efficiencies $\eta^{(\pm)}$ associated with measurement outcomes of $\mathbf{C}$ is likewise shown in Fig.~\ref{fig:optangle_ent} for complementary measurement settings.


\section{Methods and error analysis}\label{sec:metods_error}

For the proof-of-principle demonstration of the extended heat engine, we implemented its quantum circuit on the \texttt{ibmq\_nairobi} ($7$~qubits) and \texttt{ibmq\_kyoto} ($127$~qubits) processors, accessed via the IBM Quantum Experience platform~\cite{ibm_retired}. To reconstruct the joint state of the working medium $\mathbf{Q}$ and the control system $\mathbf{C}$ after each stroke, we performed complete quantum state tomography. For each experimental point, the circuit was run up to the desired stroke and executed $10$~times at different time intervals, averaging over $8000$~shots per configuration. Statistical fluctuations in the reconstructed density matrices define the error bars shown in Section~\ref{subsec:exp}.

\begin{figure}[t!]
    \centering
    \includegraphics[width=\columnwidth]{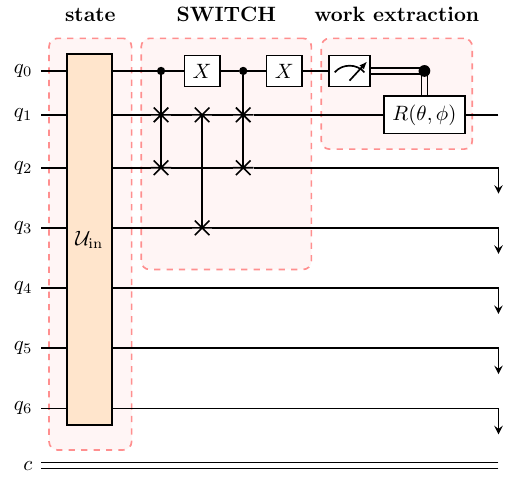}
    \caption[Complete circuit implementing the heat engine.]{\textbf{Complete circuit implementing the heat engine.} Transparent pink blocks indicate: (i) the circuit $\mathcal{U}_{\rm in}$ preparing one of the initial states shown in Fig.~\ref{fig:preparation}; (ii) the quantum SWITCH implementing the SCO between the measurement apparata $\mathbf{A}$ and $\mathbf{A}'$; and (iii) the unitary work-extraction stroke, realized through a generalized rotation~\eqref{app:eq:genRotation} that aligns the state of~$\mathbf{Q}$ along the $z$-axis of the Bloch sphere. In the depicted circuit, the quantum SWITCH operates in the coherent mode: a measurement on $q_0$ is performed, and its outcome conditions the rotation angles in the work-extraction stroke. In the incoherent mode, by contrast, the qubit $q_0$ is traced out prior to the work-extraction stroke, which then reduces to the $\mathtt{NOT}$ gate.}
    \label{fig:circuit}
\end{figure}

The initialization stage prepares one of three representative correlation regimes between the control~$\mathbf{C}$ and the working medium~$\mathbf{Q}$, as introduced in Section~\ref{subsec:theor} and illustrated in Fig.~\ref{fig:preparation}. These comprise:
\begin{description}
\item[Uncorrelated state] a product state $\varrho_{\rm in}^{\rm unc}$ with a pure control state ($\zeta = 1$);
\item[Separable state] a classically correlated state $\varrho_{\rm in}^{\rm sep}$ with perfect correlations ($\zeta_0 = 1$, $\zeta_1 = 0$);
\item[Entangled state] an entangled state $\varrho_{\rm in}^{\mathrm{qe}}$ with perfect correlations ($\zeta_0 = 1$, $\zeta_1 = 0$, $\xi_{\beta\varepsilon} = \frac{\operatorname{sech}(\beta\varepsilon)}{2}$).
\end{description}

For a given initial configuration, the circuit shown in Fig.~\ref{fig:circuit} implements strokes~\textbf{(0)--(2)} of the heat engine protocol (see Fig.~\ref{fig:protocol}). After state preparation, the quantum SWITCH is implemented following the design of Ref.~\cite{felce2021refrigeration}. The two measurement channels are implemented according to the maps $\mathcal{M}^a$ and $\mathcal{M}^{1-a}$, while controlled-$\mathtt{SWAP}$ gates condition their order on the control qubit, thereby realizing the quantum SWITCH. Tracing out the control yields an incoherent mixture of causal orders, whereas measuring the control in the complementary basis produces their coherent superposition. The latter is achieved by applying a Hadamard gate~$\mathtt{H}$ to the control and measuring it in the computational basis, which is equivalent to a measurement in the $\{\ket{+},\ket{-}\}$ basis since $\mathtt{H}\ket{0} = \ket{+}$ and $\mathtt{H}\ket{1} = \ket{-}$.

The work-extraction stroke (see Fig.~\ref{fig:circuit}) is implemented as a unitary transformation acting on $\mathbf{Q}$, ensuring that the stroke is isentropic. The corresponding unitary, associated with the measurement apparatus~$\mathbf{B}$, is implemented as a $\mathtt{NOT}$ gate in the incoherent mode of the quantum SWITCH, or whenever the post-measurement state of the working medium exhibits no coherence (i.e., $\rho_{01,\pm}^{(1)} = 0$). Otherwise, it is realized as a generalized rotation,
\begin{align}
R(\theta, \phi) &= R_y\Bigl(-\theta + \pi\frac{1-\sigma}{2}\Bigr)R_z(-\phi),\label{app:eq:genRotation}
\end{align}
where $R_y(\alpha)$ and $R_z(\alpha)$ denote rotations by an angle~$\alpha$ about the $y$- and $z$-axes, respectively. The goal of this transformation is to align the Bloch vector of~$\mathbf{Q}$ along the $z$-axis; depending on the resulting orientation, the stroke corresponds to work extraction (positive) or work investment (negative). We determine the rotation angles as
\begin{eqnarray}
\label{app:eq:alpha1} \theta &=&
\begin{cases}
\arccos\!\left(\dfrac{r_z}{|\vec r|}\right), & |\vec r|>0,\\[4pt]
0, & |\vec r|=0,
\end{cases} \\
\label{app:eq:alpha2} \phi &=&
\begin{cases}
\arg(r_x + i r_y), & r_x^2 + r_y^2 > 0,\\[4pt]
0, & r_x = r_y = 0,
\end{cases}
\end{eqnarray}
with
\begin{align}
\nonumber \vec{r} &= (r_x,\, r_y,\, r_z) \\
&= \Bigl(2\operatorname{Re}[\rho_{01, \pm}^{(1)}],\, -2\operatorname{Im}[\rho_{01, \pm}^{(1)}],\, 2\bar{a}_{\pm}-1\Bigr),
\end{align}
and $\sigma \in \{+1, -1\}$ is chosen to ensure the desired alignment along the $z$-axis ($\sigma = +1$ for work extraction, $\sigma = -1$ for work investment). Circuit optimization was handled automatically by the IBM Qiskit compiler, ensuring logical gate equivalence while minimizing total gate depth and accumulated error. The reconstructed joint states of the working medium and control were post-selected on $\ket{\pm}_{\mathrm{c}}$ and partially traced as required. Statistical uncertainties were obtained via Monte Carlo resampling of the tomographic data.

Finally, the local Gibbs state~\eqref{eq:iniGibbs} of the working medium was prepared experimentally for each value of~$a$ and for each correlation regime, as shown in Fig.~\ref{fig:preparation}. Since $\rho^{(0)}$ is independent of~$a$ and of the chosen correlation configuration, the experimental uncertainty associated with the initial temperature of~$\mathbf{Q}$ was evaluated by averaging over all regimes. No explicit noise-mitigation procedure or decoherence model was applied in our experiments; the implemented maps $\mathcal{M}^{\lambda}$ are realized through gate-based circuit decompositions using the native operations of the IBM devices, which approximate the intended CPTP channels up to hardware noise. Because internal pulse-level control is not accessible on the IBM Quantum Experience platform, we quantify these deviations operationally through the reconstructed thermal state $\rho^{(0)}$, whose uncertainty yields the effective temperature range $k_{\mathrm B}T = 1.65 \pm 0.02$. This provides the error estimate used throughout our analysis.

\bibliographystyle{apsrev4-1}
\bibliography{main}

\end{document}